\makeatletter

\font\tenmsx=msxm10
\font\sevenmsx=msxm7
\font\fivemsx=msxm5
\font\tenmsy=msym10
\font\sevenmsy=msym7
\font\fivemsy=msym5
\newfam\msxfam
\newfam\msyfam
\textfont\msxfam=\tenmsx  \scriptfont\msxfam=\sevenmsx
  \scriptscriptfont\msxfam=\fivemsx
\textfont\msyfam=\tenmsy  \scriptfont\msyfam=\sevenmsy
  \scriptscriptfont\msyfam=\fivemsy

\def\hexnumber@#1{\ifnum#1<10 \number#1\else
 \ifnum#1=10 A\else\ifnum#1=11 B\else\ifnum#1=12 C\else
 \ifnum#1=13 D\else\ifnum#1=14 E\else\ifnum#1=15 F\fi\fi\fi\fi\fi\fi\fi}

\def\msx@{\hexnumber@\msxfam}
\def\msy@{\hexnumber@\msyfam}
\mathchardef\boxdot="2\msx@00
\mathchardef\boxplus="2\msx@01
\mathchardef\boxtimes="2\msx@02
\mathchardef\square="0\msx@03
\mathchardef\blacksquare="0\msx@04
\mathchardef\centerdot="2\msx@05
\mathchardef\lozenge="0\msx@06
\mathchardef\blacklozenge="0\msx@07
\mathchardef\circlearrowright="3\msx@08
\mathchardef\circlearrowleft="3\msx@09
\mathchardef\rightleftharpoons="3\msx@0A
\mathchardef\leftrightharpoons="3\msx@0B
\mathchardef\boxminus="2\msx@0C
\mathchardef\Vdash="3\msx@0D
\mathchardef\Vvdash="3\msx@0E
\mathchardef\vDash="3\msx@0F
\mathchardef\twoheadrightarrow="3\msx@10
\mathchardef\twoheadleftarrow="3\msx@11
\mathchardef\leftleftarrows="3\msx@12
\mathchardef\rightrightarrows="3\msx@13
\mathchardef\upuparrows="3\msx@14
\mathchardef\downdownarrows="3\msx@15
\mathchardef\upharpoonright="3\msx@16

\mathchardef\downharpoonright="3\msx@17
\mathchardef\upharpoonleft="3\msx@18
\mathchardef\downharpoonleft="3\msx@19
\mathchardef\rightarrowtail="3\msx@1A
\mathchardef\leftarrowtail="3\msx@1B
\mathchardef\leftrightarrows="3\msx@1C
\mathchardef\rightleftarrows="3\msx@1D
\mathchardef\Lsh="3\msx@1E
\mathchardef\Rsh="3\msx@1F
\mathchardef\rightsquigarrow="3\msx@20
\mathchardef\leftrightsquigarrow="3\msx@21
\mathchardef\looparrowleft="3\msx@22
\mathchardef\looparrowright="3\msx@23
\mathchardef\circeq="3\msx@24
\mathchardef\succsim="3\msx@25
\mathchardef\gtrsim="3\msx@26
\mathchardef\gtrapprox="3\msx@27
\mathchardef\multimap="3\msx@28
\mathchardef\therefore="3\msx@29
\mathchardef\because="3\msx@2A
\mathchardef\doteqdot="3\msx@2B

\mathchardef\triangleq="3\msx@2C
\mathchardef\precsim="3\msx@2D
\mathchardef\lesssim="3\msx@2E
\mathchardef\lessapprox="3\msx@2F
\mathchardef\eqslantless="3\msx@30
\mathchardef\eqslantgtr="3\msx@31
\mathchardef\curlyeqprec="3\msx@32
\mathchardef\curlyeqsucc="3\msx@33
\mathchardef\preccurlyeq="3\msx@34
\mathchardef\leqq="3\msx@35
\mathchardef\leqslant="3\msx@36
\mathchardef\lessgtr="3\msx@37
\mathchardef\backprime="0\msx@38
\mathchardef\risingdotseq="3\msx@3A
\mathchardef\fallingdotseq="3\msx@3B
\mathchardef\succcurlyeq="3\msx@3C
\mathchardef\geqq="3\msx@3D
\mathchardef\geqslant="3\msx@3E
\mathchardef\gtrless="3\msx@3F
\mathchardef\sqsubset="3\msx@40
\mathchardef\sqsupset="3\msx@41
\mathchardef\trianglerighteq="3\msx@44
\mathchardef\trianglelefteq="3\msx@45
\mathchardef\bigstar="0\msx@46
\mathchardef\between="3\msx@47
\mathchardef\blacktriangledown="0\msx@48
\mathchardef\blacktriangleright="3\msx@49
\mathchardef\blacktriangleleft="3\msx@4A
\mathchardef\blacktriangle="0\msx@4E
\mathchardef\triangledown="0\msx@4F
\mathchardef\eqcirc="3\msx@50
\mathchardef\lesseqgtr="3\msx@51
\mathchardef\gtreqless="3\msx@52
\mathchardef\lesseqqgtr="3\msx@53
\mathchardef\gtreqqless="3\msx@54
\mathchardef\Rrightarrow="3\msx@56
\mathchardef\Lleftarrow="3\msx@57
\mathchardef\veebar="2\msx@59
\mathchardef\barwedge="2\msx@5A
\mathchardef\doublebarwedge="2\msx@5B
\mathchardef\angle="0\msx@5C
\mathchardef\measuredangle="0\msx@5D
\mathchardef\sphericalangle="0\msx@5E
\mathchardef\varpropto="3\msx@5F
\mathchardef\smallsmile="3\msx@60
\mathchardef\smallfrown="3\msx@61
\mathchardef\Subset="3\msx@62
\mathchardef\Supset="3\msx@63
\mathchardef\Cup="2\msx@64

\mathchardef\Cap="2\msx@65

\mathchardef\curlywedge="2\msx@66
\mathchardef\curlyvee="2\msx@67
\mathchardef\leftthreetimes="2\msx@68
\mathchardef\rightthreetimes="2\msx@69
\mathchardef\subseteqq="3\msx@6A
\mathchardef\supseteqq="3\msx@6B
\mathchardef\bumpeq="3\msx@6C
\mathchardef\Bumpeq="3\msx@6D
\mathchardef\lll="3\msx@6E

\mathchardef\ggg="3\msx@6F

\mathchardef\circledS="0\msx@73
\mathchardef\pitchfork="3\msx@74
\mathchardef\dotplus="2\msx@75
\mathchardef\backsim="3\msx@76
\mathchardef\backsimeq="3\msx@77
\mathchardef\complement="0\msx@7B
\mathchardef\intercal="2\msx@7C
\mathchardef\circledcirc="2\msx@7D
\mathchardef\circledast="2\msx@7E
\mathchardef\circleddash="2\msx@7F
\def\ulcorner{\delimiter"4\msx@70\msx@70 }
\def\urcorner{\delimiter"5\msx@71\msx@71 }
\def\llcorner{\delimiter"4\msx@78\msx@78 }
\def\lrcorner{\delimiter"5\msx@79\msx@79 }
\def\yen{\mathhexbox\msx@55 }
\def\checkmark{\mathhexbox\msx@58 }
\def\circledR{\mathhexbox\msx@72 }
\def\maltese{\mathhexbox\msx@7A }
\mathchardef\lvertneqq="3\msy@00
\mathchardef\gvertneqq="3\msy@01
\mathchardef\nleq="3\msy@02
\mathchardef\ngeq="3\msy@03
\mathchardef\nless="3\msy@04
\mathchardef\ngtr="3\msy@05
\mathchardef\nprec="3\msy@06
\mathchardef\nsucc="3\msy@07
\mathchardef\lneqq="3\msy@08
\mathchardef\gneqq="3\msy@09
\mathchardef\nleqslant="3\msy@0A
\mathchardef\ngeqslant="3\msy@0B
\mathchardef\lneq="3\msy@0C
\mathchardef\gneq="3\msy@0D
\mathchardef\npreceq="3\msy@0E
\mathchardef\nsucceq="3\msy@0F
\mathchardef\precnsim="3\msy@10
\mathchardef\succnsim="3\msy@11
\mathchardef\lnsim="3\msy@12
\mathchardef\gnsim="3\msy@13
\mathchardef\nleqq="3\msy@14
\mathchardef\ngeqq="3\msy@15
\mathchardef\precneqq="3\msy@16
\mathchardef\succneqq="3\msy@17
\mathchardef\precnapprox="3\msy@18
\mathchardef\succnapprox="3\msy@19
\mathchardef\lnapprox="3\msy@1A
\mathchardef\gnapprox="3\msy@1B
\mathchardef\nsim="3\msy@1C
\mathchardef\napprox="3\msy@1D
\mathchardef\nsubseteqq="3\msy@22
\mathchardef\nsupseteqq="3\msy@23
\mathchardef\subsetneqq="3\msy@24
\mathchardef\supsetneqq="3\msy@25
\mathchardef\subsetneq="3\msy@28
\mathchardef\supsetneq="3\msy@29
\mathchardef\nsubseteq="3\msy@2A
\mathchardef\nsupseteq="3\msy@2B
\mathchardef\nparallel="3\msy@2C
\mathchardef\nmid="3\msy@2D
\mathchardef\nshortmid="3\msy@2E
\mathchardef\nshortparallel="3\msy@2F
\mathchardef\nvdash="3\msy@30
\mathchardef\nVdash="3\msy@31
\mathchardef\nvDash="3\msy@32
\mathchardef\nVDash="3\msy@33
\mathchardef\ntrianglerighteq="3\msy@34
\mathchardef\ntrianglelefteq="3\msy@35
\mathchardef\ntriangleleft="3\msy@36
\mathchardef\ntriangleright="3\msy@37
\mathchardef\nleftarrow="3\msy@38
\mathchardef\nrightarrow="3\msy@39
\mathchardef\nLeftarrow="3\msy@3A
\mathchardef\nRightarrow="3\msy@3B
\mathchardef\nLeftrightarrow="3\msy@3C
\mathchardef\nleftrightarrow="3\msy@3D
\mathchardef\divideontimes="2\msy@3E
\mathchardef\varnothing="0\msy@3F
\mathchardef\nexists="0\msy@40
\mathchardef\mho="0\msy@66
\mathchardef\thorn="0\msy@67
\mathchardef\beth="0\msy@69
\mathchardef\gimel="0\msy@6A
\mathchardef\daleth="0\msy@6B
\mathchardef\lessdot="3\msy@6C
\mathchardef\gtrdot="3\msy@6D
\mathchardef\ltimes="2\msy@6E
\mathchardef\rtimes="2\msy@6F
\mathchardef\shortmid="3\msy@70
\mathchardef\shortparallel="3\msy@71
\mathchardef\smallsetminus="2\msy@72
\mathchardef\thicksim="3\msy@73
\mathchardef\thickapprox="3\msy@74
\mathchardef\approxeq="3\msy@75
\mathchardef\succapprox="3\msy@76
\mathchardef\precapprox="3\msy@77
\mathchardef\curvearrowleft="3\msy@78
\mathchardef\curvearrowright="3\msy@79
\mathchardef\digamma="0\msy@7A
\mathchardef\varkappa="0\msy@7B
\mathchardef\hslash="0\msy@7D
\mathchardef\hbar="0\msy@7E
\mathchardef\backepsilon="3\msy@7F
\def\Bbb{\ifmmode\let\next\Bbb@\else
 \def\next{\errmessage{Use \string\Bbb\space only in math mode}}\fi\next}
\def\Bbb@#1{{\Bbb@@{#1}}}
\def\Bbb@@#1{\fam\msyfam#1}

\def\inv{^{\raise.15ex\hbox{${
  \scriptscriptstyle -}$}\kern-.05em 1}}

\def\Dsl{\,\raise.15ex\hbox{$/$}\mkern-13.5mu D}
\def\dsl{\raise.15ex\hbox{$/$}\kern-.57em\hbox{$\partial$}}

\def\lspace{\ifx\answ\bigans{}\else\qquad\fi}

\def\CR{\hbox{{$\cal R$}}}

\def\lform{\hbox{$\sqcup$}\llap{\hbox{$\sqcap$}}}
\def\darr#1{\raise1.5ex\hbox{$\leftrightarrow$}
\mkern-16.5mu #1}
\def\h{{{1\over2}}}

\def\INT{{\textstyle \int\kern-.642em\int}}

\def\C{{\Bbb C}}
\def\Z{{\Bbb Z}}

\def\eps{{\epsilon}}

\def\cocross{{>\!\!\!\triangleleft}}

\def\tens{\mathop{\otimes}}

\def\isom{{\cong}}

\def\Ad{{\rm Ad}}

\def\image{{\rm image}\, }

\def\id{{\rm id}}

\def\Lin{{\rm Lin}}

\def\eqn#1#2{\begin{equation}#2\label{#1}\end{equation}}
\def\o{{}_{(1)}}\def\t{{}_{(2)}}\def\th{{}_{(3)}}

\def\bo{{}^{\bar{(1)}}}\def\bt{{}^{\bar{(2)}}}

\def\und#1{{\underline {#1}}}

\def\uo{{{}^{(1)}}}\def\ut{{{}^{(2)}}}

\def\text#1{\mbox{\rm #1}}
\def\note#1{}

\def\blacksquare{{\lform}}%AMS Tex Fakes
\def\frac#1#2{{{#1\over#2}}}

\def\goth#1{{#1}}

\def\proof{\goodbreak\noindent{\bf Proof\quad}}

\def\endproof{{\ $\lform$}\bigskip }

\def\align#1{\begin{eqnarray*}#1\end{eqnarray*}}
\def\und#1{{\underline{#1}}}

\def\vect{{\bf t}}\def\vecv{{\bf v}}

\def\<{\langle}
\def\>{\rangle}

\makeatother

\def\vect{{\bf t}}
\def\vecv{{\bf v}}
\def\vecw{{\bf w}}
\def\goth#1{{#1}}

\documentstyle[12pt]{article}
\begin{document}
\newtheorem{prop}{Proposition}[section]
\newtheorem{lemma}[prop]{Lemma}
\newtheorem{thm}[prop]{Theorem}
\newtheorem{df}[prop]{Definition}
\newtheorem{rk}[prop]{Remark}
\newtheorem{cor}[prop]{Corollary}
\newtheorem{ex}[prop]{Example}

\baselineskip 13pt

\markright{DAMTP/92-27/revised}

\title{Quantum group gauge theory on quantum spaces}

\author{Tomasz Brzezi\'{n}ski \thanks{Supported by St. John's College,
Cambridge \& KBN grant 2 0218 91 01} \& Shahn Majid \thanks{SERC
Fellow and Drapers Fellow of Pembroke College, Cambridge}\\ Department
of Applied Mathematics \\ \& Theoretical
Physics \\ University of Cambridge \\ CB3 9EW, U.K.}
\date{May 1992 -- revised March 1993}

\maketitle

\begin{quote}ABSTRACT We construct quantum group-valued canonical connections
on quantum homogeneous spaces, including a q-deformed Dirac monopole
on the quantum sphere of Podles quantum differential
coming from the 3-D calculus of Woronowicz on $SU_q(2)$ .
The construction is presented within the setting of a general
theory of quantum principal bundles with quantum group
(Hopf algebra) fiber, associated quantum vector bundles and connection
one-forms.
Both the base space (spacetime) and the total space are non-commutative
algebras
(quantum spaces).

\end{quote}

\baselineskip 13pt
\tableofcontents

\baselineskip 20pt

\section{Introduction}

Non-commutative geometry is based on the simple idea that in place of
working with the points on a space or
manifold $M$ we may work equivalently with the algebra $C(M)$ of
functions on $M$. In this algebraic form
we need not suppose that the algebra is commutative. A non-commutative
algebra $B$ when viewed as if it is the functions on some
(non-existing) space is called a {\em quantum space}. The process of
quantization in physics precisely
turns the commutative algebra of observables of a classical system
into a non-commutative one, hence the terminology.

Well-established in this programme are notions of integration,
differential enveloping algebras (roughly speaking, differential
forms),  cohomology classes and Chern-characters
\cite{Con:non}\cite{Con:alg}\cite{ConRie:yan}\cite{Kas:cyc}. Not only
vector bundles but also $GL(n)$ frame bundles can be understood in this context
\cite{Con:cyc}. This line of development can also be expected to have
important applications in physics, see \cite{ConLot:par}\cite{Con:met} and
also \cite{DKM:non}. An important theme in these works is the use of
non-commutative geometry to formulate some kind of generalization of gauge
theory.

In contrast to this existing approach to non-commutative geometry, we
would like to take here some steps towards developing a gauge theory in which
a more fundamental role is played by quantum groups, appearing as the fiber
of a quantum principal bundle and playing the role of structure group in the
group
of gauge transformations. Here quantum groups (Hopf algebras) are commonly
accepted as the natural analogue in non-commutative geometry of a group.
Moreover, nowadays a rich supply of true quantum groups (neither commutative
nor dual to a commutative one) are
known\cite{Dri}\cite{Wor:twi}\cite{FRT:lie}\cite{Ma:hop}.
Hence it seems an appropriate time to develop such a formalism.
Most of the formalism needed is in fact relatively straightforward
(and not incompatible with existing ideas in non-commutative geometry)
and from this point of view perhaps the most significant part of the paper is
the rich
class of examples that we also provide. These examples are modelled on the
principal bundles
and canonical connections associated to suitable homogeneous spaces. We present
the examples
and some aspects of the formal setting in which they should be viewed.

We would like to mention at least two physical motivations for
developing such a quantum-group gauge theory. The first is a formal interest in
developing
q-deformed versions of many constructions in physics. The introduction of such
a parameter $q$ may then be useful for example to regularise infinities that
arise in the corresponding
quantum field theory, which could appear now as poles in the
$q$-plane\cite{Ma:reg}.
After renormalizing (using identities from $q$-analysis) one could set $q=1$.
One may envisage other applications also in which $q$ has a more physical
meaning. The most popular quantum groups as
in \cite{Dri}\cite{FRT:lie} should be understood precisely as such
$q$-deformations
rather than arising literally from a process of physical quantization. The
differential structure
on quantum groups and certain quantum spaces are also well-understood from this
deformation
point of view and we shall need to make use of this when constructing examples.

The second and more standard motivation arises from the general indication that
the small-scale structure of space time is not well-modelled by usual continuum
geometry. At the Planck scale one may reasonably expect that our notion of
geometry has to be modified to include quantum effects also. Non-commutative
geometry clearly has the potential to do this, and this is surely one of the
long-term motivations behind some of the serious attempts to develop it, such
as \cite{Con:met}. It was also the motivation behind the introduction of the
class of Hopf algebras in \cite{Ma:hop}. These (unlike the more familiar
quantum groups) are genuinely the quantum algebras of observables of certain
quantum systems. It is hoped that some of these various constructions can
ultimately be combined with the quantum group gauge theory developed here.

An outline of the paper is as follows. In order to provide the context for
our principal bundles we shall have to introduce a significant amount of
formalism. Our preliminary Section~2 begins by recalling the standard
approach to quantum differential calculus. Given an algebra $B$ (such as the
quantum base space of the bundle) one can take as exterior algebra the
universal differential envelope
$\Omega B$ as in \cite{Con:non}\cite{Kas:cyc}. One can also construct other
differential calculi
as quotients of it. The one-forms are denoted $\Gamma_B$.

The axioms and properties of Hopf algebras are recalled in Section~3 which then
proceeds to give the most elementary version of the theory: the version in a
local co-ordinate system valid for the case of
trivial bundles. Gauge fields, curvature forms, sections, covariant
derivatives and gauge transformation properties are defined in an obvious way
that closely resembles formulae familiar to physicists for ordinary
gauge fields. This section is also preliminary and serves to introduce several
standard notions that will play an important role
in the later sections, such as coactions, comodule algebras and the convolution
product $*$.
It also provides the local picture to which we feel any reasonable theory of
principal and associated bundles
should reduce in the trivial case. An unusual feature encountered here even at
the level of trivial
bundles, is that the group of gauge transformations (which remains an ordinary
group) does not
consist only of algebra maps from $A$ (the quantum group) to $B$ (the base
quantum space) as one might naively expect, but needs to be enlarged as soon as
$B$ is non-commutative.

In Section~4 we pass to the more abstract setting needed to handle
non-trivial bundles. By definition these are algebras $P$ (the total
quantum space) on which $A$ coacts with fixed-point subalgebra $B$. In
addition, we need some condition corresponding
to freeness of the action and an exactness condition to replace smoothness and
dimension arguments in the classical situation. We do this in terms of a map
$\widetilde{\
}$
generating the fundamental vector fields on $P$ induced by the coaction of
$A$. One can also formulate the `local triviality'
of the situation in terms of the patching together of
a collection of trivial bundles related to each other by gauge
transformations. Other `purely quantum' possibilities also
open up once the algebras are non-commutative.

In this abstract setting one works with a connection as a splitting of the
tangent or cotangent
space (in our case it is convenient to use the latter). A main (if tedious)
task in any textbook on differential
geometry is to relate this abstract definition of a connection to another
definition as a connection one-form on
$P$, and to show in the trivial case that this in turn implies the usual local
picture of gauge fields relative to a choice of trivialization. This is the
main result on Section~4. The general theory is further continued for
associated vector bundle in Appendix A. Although relatively straightforward,
there are a number of subtleties arising from the non-commutativity of the
algebras and our propositions clarify and justify the various choices that are
needed.

Since many readers may not be familiar with the necessary background in quantum
differential calculi, we begin in Section~4 with the most accessible case
of the universal differential envelope $\Omega P$. We then come
in the second half of the section to the non-universal calculi. We do not
wish to claim that our formulation is the last word on this topic, but it is
one that
is general enough to include our current range of examples. It not only
provides some kind of setting for the examples, but also provides for their
local description via the propositions in this section and in Appendix A.

Finally we are in a position in Section~5 to construct our examples of quantum
principal bundles
and connections on them, based on
quantum homogeneous spaces and their canonical connections.
By quantum homogeneous space we mean a pair of
quantum groups $P\to A$ (where the Hopf algebra surjection corresponds
to the inclusion of the structure group as a subgroup in the classical case)
subject to certain
conditions. For a
connection one needs in the classical case that the subgroup is
reductive -- the analogue of this for our purposes is that we
need to split the surjection by an Ad-covariant algebra map
$i:A\hookrightarrow P$ at least locally.

The simplest non-trivial case is then examined in detail, with $A=k(S^1)$
and $P=SO_q(3)$. Here the base
is the quantum
sphere of Podle\'s\cite{Pod:sph} and the bundle is a quantum monopole
bundle. The canonical connection
is studied, and with the correct quantum-differential calculus (not the
universal one) it
recovers the standard U(1)-Dirac
monopole in the limit $q\to 1$. The differential calculus chosen for this
example is inherited from
the 3-D one on $SU_q(2)$ introduced in \cite{Wor:dif}. It demonstrates the
usefulness of the various conditions and results of the general theory of
Section~4,
and also connects
ultimately with a local description as in Section~3.

Finally, because our formulae for abstract Hopf algebras
may be a little unfamiliar, we collect together in Appendix B the
various formulae in the case when $A$ is a matrix quantum group. Here
the convolution product $*$ corresponds to matrix multiplication.

Throughout the paper our algebras are assumed  unital algebras over a
field $k$ of characteristic not 2. It is
hoped that our algebraic formulation
may be useful in purely algebraic work also, such as the introduction
of new invariants of algebras
and Hopf algebras based on gauge theory. In the other direction, the
algebraic setting may be useful
even in the classical case in the form of finite models of gauge
theory -- comparable to finite lattice
models of gauge theory but preserving much more of the geometrical
picture in an exact form. For example, the space
of gauge fields relative to a given one could be some
finite-dimensional space which could then be integrated over.
For infinite systems of course one needs to work with operator
algebras. Here we would like to note that all
our constructions are fully compatible with $*$-algebra structures
placed on the algebras, and hence suitable
for such a treatment. We
will, however, have enough to do in the present paper at a purely
algebraic level.

\section{Preliminaries about universal differential calculus}

Here we recall some standard facts about differential calculus on an algebra.
We refer to  \cite{Con:non}\cite{Kas:cyc} for further details.

The general notion is that of a {\em $\Z_{2}$-graded differential algebra},
meaning
an algebra $\Xi$ equipped with ${\bf
Z}_{2}$-grading (denoted by $\partial$) and a linear operation $d :
\Xi \rightarrow \Xi$ of
degree 1, obeying the graded Leibniz rule and such that $d^{2} =
0$. We will say that $(\Gamma ,d)$ is a {\it first order
differential calculus} over an algebra $A$ if $d: { A} \rightarrow
\Gamma$ is a linear map obeying the Leibniz rule, $\Gamma$ is a bimodule over
$A$ and every element of $\Gamma$ is of the form $\sum_{k=1}
a_{k}db_{k}$, where $a_{k}, b_{k} \in A$.
To every first order differential calculus $(\Gamma, d)$ over $A$
one can associate a $\Z_2$-graded differential algebra $(\Omega (A),d)$ in the
following way. Firstly, one defines $\Omega^{0}({A}) = A$ and
$$\Omega^{n}(A) \subset \Gamma \otimes_{A} \Gamma \otimes
_{A} \cdots \otimes_{A} \Gamma = \Gamma ^{\otimes _A n}$$
for $n >0$, as a set spanned by all elements:
\begin{equation}
(a_{0},a_{1}, \ldots ,a_{n}) = a_{0} \otimes_{A} da_{1}
\otimes_{A} \cdots \otimes_{A} da_{n}
\end{equation}
for any $a_{0},a_{1}, \ldots , a_{n} \in A$. One can then introduce the natural
$\Z_{2}$-grading, $\partial \omega_{n} = n({\rm mod} \; 2)$ and
 define $\Omega ({A}) = \bigoplus_{n=1}^{\infty} \Omega
^{n} ({A})$. The product of $(a_{0}, \ldots , a_{n}) \in \Omega
^{n} ({A})$ and $(a_{n+1}, \ldots , a_{n+m}) \in \Omega
^{m-1} ({A})$ is given by
\begin{equation}
(a_{0}, \ldots ,a_{n})(a_{n+1}, \ldots ,a_{m+n}) = \sum
_{i=0}^{n} (-1)^{i} (a_{0}, \ldots, a_{n-1-i}, a_{n-i}a_{n-i+1},
a_{n-i+2}, \ldots, a_{n+m})
\label{modulo}
\end{equation}
and $d$ is extended to the whole of $\Omega ({A})$ by:
$$d(a_{0}, a_{1}, \ldots ,a_{n}) = (1, a_{0}, a_{1}, \ldots , a_{n})$$
$$d(1,a_{0}, a_{1}, \ldots ,a_{n}) = 0$$

$\Omega (A)$ is therefore a free tensor algebra modulo relation
(\ref{modulo}). In some cases however one can consider ideals $I^n
\subset \Omega ^n(A)$  and define the exterior algebra of $A$
associated to $\Gamma_A$ by taking quotients $\Omega^n(A) / I^n$.
Ideals $I^n$ has to be compatible with the action of the
differential $d$. In what follows we do not stress difference between
$\Omega (A)$ and suitable quotients of it.

It is known that every first order differential calculus on an algebra
${A}$ can be obtained as the quotient of a universal differential
calculus $({A}^{2}, d)$.  Here ${A} ^{2} = {\rm ker} \cdot$ (where
$\cdot : {A}
 \otimes {A} \rightarrow {A}$ is the multiplication map in
$A$) and  $d: A \rightarrow A ^{2}$ is defined by
\begin{equation}
d a = 1 \otimes a - a \otimes 1.
\end{equation}
This map $d$ clearly obeys the Leibniz rule provided ${A} ^{2}$ has the $A$
bimodule structure given by
\begin{eqnarray}
c(\sum_{k} a_{k} \otimes b_{k}) = \sum_{k} ca_{k} \otimes b_{k} \\
(\sum_{k} a_{k} \otimes b_{k} )c = \sum_{k} a_{k} \otimes b_{k}c
\end{eqnarray}
for any $\sum_{k} a_{k} \otimes b_{k} \in {A}^{2}$, $c \in A$.
Furthermore, it is easy to see that every element of ${A}^{2}$ can be
represented in the form
$\sum_{k} a_{k} db_{k}$. In this way $({A}^{2} , d)$ is indeed a first order
differential calculus
over $A$ as stated. The $\Z_2$-graded differential algebra defined by $(A^{2},
d)$  will be denoted by $(\Omega {A}, d)$  and called the {\em
differential envelope of $A$} (cf \cite{Kas:cyc}). We have the
following universality principle:
\begin{prop}
(\cite{Kas:cyc}, \cite{CoqKas:rem}) Let $(\Xi ,\delta)$  be any
differential algebra with unity, and $A$ any
algebra with unity. Any 0-degree homomorphism $\alpha: {A} \rightarrow
\Xi$ can be lifted to a unique 0-degree homomorphism $\theta : \Omega
{A} \rightarrow \Xi$ such that $\theta \mid_{A} = \alpha$
and $\theta\circ d = \delta \circ \alpha$.
\label{prop.universal}
\end{prop}

By the natural identification $A \otimes_A A \cong A$  one can easily
prove by induction (see
\cite{Con:non}) that
\[ \Omega ^nA = \{ \rho \in A \otimes_k \cdots \otimes_k
A = A ^{\otimes n+1} : \forall i \in \{1, \ldots ,n\}, \cdot_i \rho = 0
\}\]
 where
$$\cdot_i = id \otimes_k id \otimes_k \cdots \otimes
_k \cdot \otimes_k \cdots \otimes_k id$$
(multiplication $\cdot$ acting in the $i, i+1$-th place). Hence
$\Omega ^n A \subset
A^{\otimes_kn+1}$. Notice that the description of $\Omega ^n A$ is
purely algebraic (i.e. it depends only on the properties of
the multiplication in $A$). In particular, this means that if $B$ is a
subalgebra of $A$ with $j: B \hookrightarrow A$ the inclusion map,
then $j$ can be extended as an inclusion $j: \Omega B
\hookrightarrow \Omega A$.

Proposition~\ref{prop.universal} allows one to reconstruct any
differential algebra $\Omega (A)$ as
\[ \Omega^n(A)=\Omega^n A/N^n\]
where $N^n \subset \Omega^n A$ are ideals, $n=1,2,\ldots$. If
$B\subset A$ then we
will take differential structure $\Omega(B)$ as
defined by the ideals $N_B^n = N^n\cap \Omega^nB$. This assumption
implies that the inclusion $j:B\hookrightarrow A$ extends to an
inclusion $j:\Omega(B)\hookrightarrow\Omega(P)$, commuting with $d$.

\section{Gauge fields on trivial quantum vector bundles}

In this second preliminary section we present the construction of trivial
quantum vector bundles and gauge fields on them. This also serves to
introduce the basic facts and constructions for Hopf algebras (quantum groups)
which
will be needed later. The role of the
structure group is played by the
quantum group or Hopf algebra and the roles of the base and fiber are played by
algebras which can also be non-commutative (i.e. quantum spaces). In
fact the definitions presented here are a special case of a general theory
of quantum vector bundles which will be described later.
Here we would like to emphasise instead the definition of quantum
vector bundles from the point of view of gauge transformations. This gives a
self-contained picture in which all fields live on the base. This
point of view is closely related to physics and has proven to be very
fruitful. Moreover, it provides the basic local theory to which our general
abstract must reduce in the
trivial case.

Let us recall that a Hopf algebra is an associative algebra $A$ with unit
equipped with a compatible coalgebra
structure. This consists of algebra maps
 $\Delta : {A} \rightarrow {A} \otimes {A}$ (the comultiplication), $\epsilon :
{A} \rightarrow  k$ (the counit)
and a linear map $S:
{A} \rightarrow A$ (the antipode) obeying the following axioms
\begin{enumerate}
\item  $(\Delta \otimes id) \Delta = (id \otimes \Delta) \Delta$
\item  $(\epsilon \otimes id) \Delta = (id
\otimes \epsilon )\Delta = id$
\item  $ \cdot (S \otimes id) \Delta = \cdot (id
\otimes S) \Delta = \eta \circ \epsilon$.
\end{enumerate}
Here $\cdot $ denotes multiplication in $A$ and $\eta :k
\rightarrow A$ is the unit map, i.e. $\eta (\lambda ) = \lambda 1_A, \;
\forall \lambda \in k$. We adopt Sweedler's sigma notation\cite{Swe:hop},
namely
$\Delta (a) = \sum a_{(1)} \otimes a_{(2)}$, for any $a \in A$.

If $A$ is a Hopf algebra then we say that a vector space $V$ is a {\em left
$A$-comodule} if there exists a map
$\rho_{L} : V \rightarrow A \otimes V$ (a left coaction of $A$ on $V$)
such that
$$(\Delta \otimes id) \rho_{L} = (id \otimes \rho_{L}) \rho_L ,
\hspace{1cm} (\epsilon \otimes id)\rho_{L} = id $$
If $V$ is an algebra and $\rho_{L}$ is an algebra map, i.e.
$$\rho_{L}(ab) = \rho_{L}(a) \rho_{L}(b), \hspace{1cm} \rho
_{L}(1_V) = 1_{A} \otimes 1_{V} $$
then we will say that $V$ is a {\em left $A$-comodule algebra}. We will
sometimes use the explicit notation $\rho_{L} (v) = \sum v^{(\overline
1)} \otimes
v^{(\overline 2)}$ for any $v \in V$

Similarly we say that a vector space $V$ is a {\em right $A$-comodule}
if there exists a linear map
$\rho_{R} : V \rightarrow V \otimes A$ (a right coaction $A$ on $V$) such that
$$(\rho_{R} \otimes id) \rho_{R} = (id \otimes \Delta)\rho
_{R}, \hspace{1cm} (id \otimes \epsilon )\rho_{R} = id$$
If $V$ is an algebra and $\rho_{R}$ is an algebra map then we say
that $V$ is a {\em right $A$-comodule algebra}.

Given a bialgebra $A$ there is an opposite bialgebra $A^{\rm op}$ consisting of
$A$ with the opposite product.
If $A$ is a Hopf algebra with bijective antipode then $S^{-1}$ makes $A^{\rm
op}$ also into a
Hopf algebra. When we come to the abstract theory of associated vector bundles
we will need both $A$-comodule algebras and $A^{\rm op}$-comodule algebras in
order to make a quotient tensor product algebra by the coaction (a cotensor
product).

To complete our preliminary remarks on Hopf algebras we recall
the convolution product of linear maps on a Hopf algebra (or
coalgebra) $A$. Let $B$
be an algebra and $f_{1}, f_{2}: A \rightarrow B$ two linear maps. The
{\em convolution product} of
 $f_{1}$ and $f_{2}$ (denoted by $g = f_{1} * f_{2}$) is
the linear map $g: A \rightarrow   B$ given by $g(a) =
\sum f_{1}(a_{(1)}) f_{2}(a_{(2)})$ for any $a \in A$. The convolution
product is associative and makes the set $\Lin(A,B)$ into an algebra.
Note that if
$B$ has a unit $\eta_{B}$ (viewed as a map) then $f * (\eta_{B}
\circ \epsilon) = (\eta_{B} \circ \epsilon) * f = f$, so that $\eta
_{B} \circ \epsilon$ is the identity in the convolution algebra $\Lin
(A,B)$. We say that a linear map $f : A \rightarrow B$ is {\em
convolution invertible} if there exists a map $f^{-1} : A \rightarrow B$
such that $f^{-1} * f = f * f^{-1} = \eta_{B} \circ \epsilon$.
Similarly if $V$ is a left $A$-comodule and $f_{1}: A \rightarrow B$,
$f_{2} : V \rightarrow B$, then $(f_{1} * f_{2}) (v) = \sum
f_{1}(v^{(\overline 1)}) f_{2}(v^{(\overline 2)})$ for any $v \in V$.
Finally if $\Gamma$ is any bimodule of B and $f_{1}: A
\rightarrow B$, $f_{2}: V \rightarrow \Gamma$ we define $(f_{1} *
f_{2})(v) = \sum f_{1}(v^{(\overline 1)}) f_{2} (v^{(\overline 2)})$.

Now we are in a position to introduce the notion of a trivial (left)
quantum vector bundle.

\begin{df}
Let $(A, \Delta, \epsilon , S)$ be a Hopf algebra. We say that
$E(B,V,A)$ is a {\em trivial (left) quantum vector bundle}
with  base $B$, fibre $V$ and structure group $A$ if:
\begin{enumerate}
  \item $B$ is an algebra with unity;
  \item $(V,\rho_{L})$ is a left $A$-comodule algebra;
  \item $E = V \otimes B$.
\end{enumerate}
\end{df}

 Let us note that $E$ is a left $A$-comodule algebra. The coaction
$\Delta_{L} : E \rightarrow A \otimes E$ is given by $\Delta
_{L} = \rho_{L} \otimes id$ and the multiplication $(v_{1} \otimes
b_{1})(v_{2} \otimes b_{2}) = v_{1}v_{2} \otimes b_{1}b_{2}$ is the
tensor product one.

A {\em quantum gauge transformation} of our trivial vector bundle
$E(B,V,A)$ is then a convolution invertible map $\gamma : A \rightarrow B$
such that $\gamma (1) = 1$. We say that $\sigma : V \rightarrow B$
is a {\em section} of $E$ if it transforms under
the action of gauge
transformation $\gamma$ according to the law $\sigma  \buildrel
\gamma \over \longmapsto \sigma ^{\gamma} = \gamma * \sigma$. $A$ acts
on $V$ according to the left coaction $\rho_{L}$. The set of
sections of $E$ will be denoted by $\Gamma (E)$. If $\Omega (B)$ is a
differential algebra over $B$ then we also consider n-form sections
$\Gamma^n(E)$, the set of maps $V \rightarrow \Omega ^n(B)$.

To make these definitions more transparent let us consider their
classical limit (see e.g. \cite{GocSch:dif}). Let $U$ be an open set
on the base, $G$ a Lie
group, and suppose the vector space ${\Bbb C}^n$ forms a representation
of $G$. We can think
of $G$ concretely as a matrix group contained in $GL(n)$ and define,
$A = C^\infty(G)$, $V= C^\infty(\Bbb C^n)$ and $B= C^\infty(U)$. In
a suitable algebraic context, $A$ becomes a Hopf algebra and the
algebra of functions on the trivial vector bundle $E=C^\infty(\Bbb
C^n\times U)$ becomes  $V \otimes B.$ A section on the bundle
$\Bbb C^n\times U$ is a vector valued function $s: U \rightarrow
\Bbb C^n$ and a gauge transformation is a
matrix valued function $g : U \rightarrow G$. Sections and gauge
transformations give rise to algebra maps $\sigma :C^\infty(\Bbb
C^n) = V \rightarrow
B = C^\infty(U)$ and $\gamma: C^\infty(G) = A \rightarrow B =
C^\infty(U)$ respectively, induced by pull-back. Moreover
the gauge transformation $g$ acting pointwise induces a
transformation of sections $s \mapsto s^{g}$, which in components reads:
$$(s^{g})^{i} (x) = g^{i}_{j}(x) s^{j} (x)$$
for all $x \in U$. This in turn gives rise to the transformation of $\sigma$,
namely as
$$\sigma ^{\gamma} (v^{i}) = \gamma (g^{i}_{j}) \sigma (v^{j}).$$
This explains our definition of quantum gauge transformations and
sections of quantum vector bundles.

The next step in the construction of quantum-group gauge theory consists of
the definition of a covariant exterior derivative. To this end let us
assume that $(\Gamma_{B} ,d )$ is a first order differential calculus
over $B$ and $\Omega (B)$ is the differential algebra induced by it. We
say that a linear map $\nabla : \Gamma (E) \rightarrow \Gamma
^{1}(E)$ is a {\em
covariant exterior derivative} on the trivial quantum vector bundle
$E$ if for any  quantum gauge
transformation $\gamma$ on $E$, there exists map $\nabla ^{\gamma} :
\Gamma (E) \rightarrow \Gamma^1(E)$ such that for any section $\sigma
\in \Gamma (E)$,
\begin{equation}
\nabla ^{\gamma} \sigma ^{\gamma} = \gamma * (\nabla \sigma)
\label{cov.der.1}
\end{equation}
In other words, $\nabla :\Gamma (E) \rightarrow \Gamma^1 (E)$ is a covariant
exterior derivative on $E$ if $\nabla$ transforms under a gauge
transformation $\gamma$ according to the rule:
\begin{equation}
 \nabla \buildrel \gamma \over \longmapsto \nabla ^{\gamma} = \gamma
* \nabla  \gamma ^{-1} *
\label{cov.der.2}
\end{equation}

Just as in the classical case we have the following:
\begin{prop}
Let $E(B,V,A)$ be a trivial quantum bundle. If
a map $\beta : A \rightarrow \Gamma_{B}$ transforms by the quantum gauge
transformation $\gamma$ of $E$ as
\begin{equation}
\beta \buildrel \gamma \over \longmapsto \beta ^{\gamma} = \gamma *
\beta * \gamma ^{-1} + \gamma * d(\gamma ^{-1})
\label{connection.1}
\end{equation}
then the map $\nabla : \Gamma (E) \rightarrow \Gamma^1(E)$ given by
\begin{equation}
\nabla = d + \beta *
\label{cov.der.3}
\end{equation}
is a covariant exterior derivative on $E$.
\label{prop.cov.der}
\end{prop}
\proof
We have to check that the linear operation $\nabla$ given by equation
(\ref{cov.der.3})  transforms according to the rule
(\ref{cov.der.1}). For any section $\sigma \in \Gamma (E)$ we have
\begin{eqnarray*}
\nabla ^{\gamma} \sigma ^{\gamma} & = & d \sigma ^{\gamma} + \beta
^{\gamma} * \sigma ^{\gamma} = d(\gamma * \sigma) + (\gamma * \beta *
\gamma ^{-1} + \gamma * d (\gamma ^{-1}))* \gamma * \sigma\\
& = & d \gamma * \sigma + \gamma * d \sigma + \gamma * \beta * \sigma
- d\gamma * \sigma = \gamma * (\nabla \sigma).
\end{eqnarray*}
Hence $\nabla$ transforms as a covariant derivative and
the result follows. \endproof

A map $\beta : A \rightarrow \Gamma_{B}$ as in Proposition
\ref{prop.cov.der} is called a {\em connection one-form} on $E$  or simply
a connection on $E$ (or quantum gauge field). The transformation rule for
connections implies the following :
\begin{prop}
Let $\gamma , \gamma ' : A \rightarrow B$ be two gauge transformations
on the trivial quantum vector bundle $E(B,V,A)$ and let $\beta : A
\rightarrow \Gamma_{B}$ be a connection on $E$. Then
\begin{equation}
(\beta ^{\gamma})^{\gamma '} = \beta ^{\gamma ' * \gamma}.
\end{equation}
\end{prop}
{\bf Proof} The proof is based on direct use of the rule
(\ref{connection.1}),
namely
\begin{eqnarray*}
(\beta ^{\gamma})^{\gamma '} & = & \gamma ' * \beta ^{\gamma} *
(\gamma ')^{-1} + \gamma ' * d ({\gamma '} ^{-1}) \\
& = & \gamma ' * \gamma * \beta * \gamma ^{-1} * {\gamma '} ^{-1} +
\gamma ' * \gamma * d(\gamma ^{-1}) * {\gamma '} ^{-1} + \gamma ' *
d({\gamma '} ^{-1}) \\
& = & \gamma ' * \gamma * \beta * \gamma ^{-1} * {\gamma '} ^{-1} +
\gamma '* d((\gamma ' * \gamma) ^{-1}) - \gamma ' * d ({\gamma '} ^{-1})
+ \gamma ' * d({\gamma '} ^{-1} ) \\
& = & \beta ^{\gamma ' * \gamma} .
\end{eqnarray*}
\endproof

To any connection $\beta$ on a trivial quantum vector bundle $E(B,V,A)$
one can associate its {\em curvature} $F:A
\rightarrow \Omega ^{2} (B)$ defined as
\begin{equation}
F = d\beta + \beta * \beta
\label{curvature.1}
\end{equation}
\begin{prop}
Let $E(B,V,A)$  be  a trivial quantum vector bundle. Let $\beta : A
\rightarrow \Gamma_{B}$ be a connection one-form on
$E$ and $F: A \rightarrow \Omega^{2} (B)$ its curvature. Then we have:
\begin{enumerate}
\item For any section $\sigma \in \Gamma (E)$
\begin{equation}
\nabla ^{2} \sigma = F * \sigma .
\label{curvature.2}
\end{equation}
\item For any quantum gauge transformation $\gamma$ of $E$
\begin{equation}
F^{\gamma} = \gamma * F * \gamma ^{-1}.
\label{curvature.3}
\end{equation}
\item The Bianchi identity
 \begin{equation}
dF +\beta *F - F * \beta = 0.
\label{bianchi}
\end{equation}
\end{enumerate}
\end{prop}

\proof
\begin{enumerate}
\item We have
\begin{eqnarray*}
\nabla ^{2} \sigma & = & \nabla (d\sigma + \beta * \sigma ) = d(\beta
* \sigma) + \beta * d \sigma + \beta * \beta * \sigma \\
& = & (d \beta + \beta *\beta) * \sigma = F * \sigma.
\end{eqnarray*}
\item The transformation law for curvature follows immediately from the
definition (\ref{curvature.1}).
\item We compute:
\begin{eqnarray*}
dF & = & d\beta * \beta - \beta * d \beta = d \beta * \beta + \beta *
\beta * \beta - \beta * \beta * \beta - \beta * d \beta \\
& = & F * \beta - \beta * F.
\end{eqnarray*}
\end{enumerate}
\endproof

As we can see, all the results obtained here are very similar to the
classical ones except that the usual product of functions is replaced by
the convolution product. In fact the convolution product appears also in the
classical construction where it corresponds to group multiplication or
the action of the
group  -- but now instead of considering groups and representations
spaces we consider
algebras of functions on them. The main difference between classical
and quantum vector bundles lies in the fact that if $E$ is a
noncommutative algebra and $A$ is a quantum group, they cannot be
interpreted as algebras of functions on an actual vector bundle and its
structure group respectively.

In the construction above we have restricted ourselves to the consideration
of left quantum vector bundles and structures related to them. But
there is well established symmetry between left and right
constructions. To conclude this section we summarize  a version of the
above results
based on right quantum vector bundles.

\begin{df}
Let $(A, \Delta, \epsilon , S)$ be a Hopf algebra. We say that
 $E(B,V,A)$ is a {\em trivial (right) quantum vector bundle}
with base $B$, fibre $V$ and structure quantum group $A$ if:
\begin{enumerate}
  \item $B$ is an algebra with unity;
  \item $(V,\rho_{R})$ is a right $A^{\rm op}$-comodule algebra;
  \item $E = B \otimes V$.
\end{enumerate}
\end{df}

Then we have the following. The induced right coaction $\Delta_{R} : E
\rightarrow E
\otimes A$ of $A$ on $E$ is given by:
$$\Delta_{R} = id \otimes \rho_{R}.$$
The gauge transformation of sections:
\begin{equation}
\sigma ^{\gamma} = \sigma * \gamma .
\label{trivial.sigma.right}
\end{equation}
The gauge transformation of covariant derivatives:
$$\nabla ^{\gamma} \sigma ^{\gamma} = (\nabla \sigma) *\gamma . $$
The gauge transformation of connection 1-forms $\beta$:
$$\beta ^{\gamma} = \gamma ^{-1} * \beta * \gamma + \gamma ^{-1} * d
\gamma .$$
Hence the covariant derivative acts on sections $\sigma \in \Gamma
(E)$, as:
\begin{equation}
\nabla \sigma = d \sigma - \sigma * \beta,
\label{trivial.nabla.1}
\end{equation}
and on the linear maps $\rho \in \Gamma^n(E)$:
\begin{equation}
 \nabla \rho = d\rho - (-1)^{n} \rho * \beta .
\label{trivial.nabla.2}
\end{equation}
Some properties of the curvature 2-form $F = d\beta + \beta * \beta$ are:
$$\nabla ^{2} \sigma = - \sigma * F$$
$$F ^{\gamma} = \gamma ^{-1} * F * \gamma$$
and the Bianchi identity :
$$dF + \beta * F - F * \beta = 0.$$

Some of the relations above need more explanation. Although they look
a little bit unusual, one can show that in fact the right-covariant
construction provides the correct classical limit (as we will see in the
next section). There are two facts which play a crucial role in this
identification. First of all let us state the following elementary lemma:
\begin{lemma}
Let $A$ be a Hopf algebra and let $(V,\rho_{R})$ be a right
$A^{\rm op}$-comodule algebra. Then $V$ is the left
$A$-comodule algebra $(V, \rho_L)$ with coaction given by
\[ \rho_{L} = \tau (id \otimes S) \rho_R \]
where $\tau$ is the usual twist map.
\end{lemma}
\proof This is an elementary exercise from the definitions above and the fact
that for any
Hopf algebra the antipode $S:A\to A$ is
an antialgebra and anticoalgebra map. \endproof

Classically, a connection 1-form $\beta$ is a Lie algebra-valued
1-form on the base. Here the Lie algebra is that of the classical gauge group
$G$. We can view it as a subset of its universal enveloping Hopf
algebra, and on this subset the
antipode acts by $-1$. In our dual picture it means that in the classical limit
we have $\beta\circ S=-\beta$ where $S$ is the antipode on $A$. Thus
if we convert our right
$A^{\rm op}$-comodule algebra to a left $A$-comodule
algebra by means of the above lemma (as is usually done) the ``-'' sign
in (\ref{trivial.nabla.1}) will be absorbed.
This is why no ``-'' sign appears in the usual classical formulae for covariant
derivatives. For general Hopf algebras the action of $S$ is more
complicated and
this cancellation is not possible. Secondly, in the classical case
the exterior algebra is graded-commutative so that $\beta$ in equation
(\ref{trivial.nabla.2}) can be written on the left of $\rho$, cancelling
the factor depending on its degree. Again, this is not possible for
a general quantum differential calculus. We note that the $(-1)^n$ is
in any case an artifact of our
writing $d$ and $\nabla$ acting from the left when, in our
right-handed conventions, they act more
simply from the right.

\section{Quantum principal bundles and connections on them}

In this section we give a general theory of quantum principal bundles. We first
work in the universal differential envelope, and come to the case of a general
differential calculus in the second subsection.

We begin with a brief outline of the classical theory of connections and fibre
bundles, following
\cite{KobNom:fou} and emphasising the aspects that we shall generalise to the
quantum case.
Let $M$ be a smooth manifold and $G$ a Lie
group. A principal bundle over $M$ consists of a smooth manifold $P$ and a
smooth
action of $G$ on $P$ such that $G$ acts freely on $P$ from the right,
i.e. $P \times G \ni (u,a)  \mapsto ua = R_{a} u \in P$ is an action and
\eqn{classical.freeness}{P\times G\to P\times P, \qquad (u,a)\mapsto (u,ua)}
is an inclusion (freeness). Moreover, $M \isom P/G$
and the canonical projection $\pi :P\rightarrow M$ is a smooth map. We
denote the principal bundle by $P(M,G)$ or simply by $P$. Locally
$P \cong M \times G$. This means that if $U \subset M$ is an open set
covered by one chart, then there exists a map $\phi_{U} : \pi ^{-1}
(U)  \rightarrow G$ such that $\phi_{U} (ua) = \phi_{U} (u)
a$ and such that the map $\pi^{-1}(U) \rightarrow U \times G$,
defined by $u \mapsto (\pi
(u) , \phi_{U} (u))$ is an isomorphism.

For each $u \in P$ let $T_{u} P$ be the tangent space of $P$ at $u$ and
$G_{u}$ the subspace of $T_{u}P$ consisting of vectors tangent to the
fibre through $u$. A connection $\Pi$ in $P$ is an assignment of a
subspace $Q_{u}$ of $T_{u}P$ to each $u \in P$ such that
\begin{equation}
T_{u}P = G_{u} \oplus Q_{u}
\label{hor.vert}
\end{equation}
 and $Q_{ua} = (R_{a})_{*} Q_{u}$ for any $u \in P$ and
$a \in G$. Here $R_{a}$ is the transformation of $P$ induced by $a \in
G$, i.e. $R_{a} u = ua$. We call $G_{u}$ the vertical subspace and
$Q_{u}$ the horizontal subspace of $T_{u}P$. Given a connection $\Pi$ in
$P$ we define a 1-form $\omega$ on $P$ with values in the Lie algebra
$g$ of $G$ in the following way. Any $\xi \in g$ induces a {\em fundamental
vector field} $\widetilde {\xi}$ on $P$. Its value on a 1-form $df$ is
\eqn{classical.tilde}{<\widetilde\xi,df>(u)={d\over dt}|_0 f(u\exp{t\xi})}
i.e. it is the differential of the right action of $G$. Now for each
$X\in T_{u}P$ we define $\omega (X)$ to
be the unique $\xi \in g$ such
that $\widetilde {\xi}$ is equal to the vertical component of $X$. Clearly
$\omega (X) = 0$ if and only if $X \in Q_{u}$.

Equivalently the connection 1-form $\omega$ is a $g$-valued 1-form on
$P$ such that $\omega (\widetilde {\xi}) = \xi$ for any $\xi \in g$ and
$(R_{a})^{*} \omega = ad (a^{-1}) \omega$, i.e. $\omega ((R_{a})_{*} X)
= ad (a^{-1}) \omega (X)$ for any $a\in g$ and any vector field $X$.
Here $ad$ denotes the adjoint representation of $G$ in $g$. Given a
connection 1-form the
corresponding projection is recovered by $\Pi=\widetilde{\  } \circ \omega$.

\subsection{The Case of Universal Differential Calculus}

We now come to the quantum (non-commutative) case. The first ingredient is an
algebra $P$ analogous to
the functions on the total space of the principal bundle. We require this to be
a comodule algebra for a Hopf algebra $A$ with right coaction $\Delta_R:P\to
P\tens A$. We assume that the action is free in the sense that the induced map
$P\tens P\to
P\tens A$ is a surjection. This is just the straightforward dualization of
(\ref{classical.freeness}) and is quite standard, see for example
\cite{Sch:pri}. We take the invariant subalgebra $B=P^A=\{u\in
P|\Delta_R(u)=u\tens 1\}$ for
the algebra analogous to the functions on the base manifold.  This is a
subalgebra
for if $u,v \in B$ then
$$\Delta_{R} (uv) = \Delta_{R} (u) \Delta_{R} (v) = (u \otimes 1)
(v \otimes 1) = (uv) \otimes 1.$$
Hence $uv \in B$. There is a natural inclusion $j : B \hookrightarrow
P$ which corresponds to the canonical projection $\pi$ in the
classical case.

Next, in place of working with tangent bundles etc, we work with forms. These
serve also to specify the differential structure on $P$ as recalled in
Section~2. For now we develop the theory only with
the differential structure given by the universal envelope $\Omega P$. The
necessary modifications for a general differential calculus will be given
later. In the case of the universal envelope our right coaction $\Delta_R$
automatically extends to $\Omega P$ as a right $A$-comodule $\Delta_{R} :
\Omega P \rightarrow \Omega P
\otimes A$. One says that the differential calculus is {\em covariant} (cf.
\cite{WesZum:cov}). Explicitly, the coaction is given here by:
\begin{equation}
\Delta_{R} (u_{0}du_{1} \cdots du_{n}) = \sum u_{0}^{(\overline 1)}
du_{1}^{(\overline 1)} \cdots du_{n}^{(\overline 1)} \otimes
u_{0}^{(\overline 2)} u_{1}^{(\overline 2)} \cdots u_{n}^{(\overline
2)}
\label{covariance}
\end{equation}
where $u_{0}, \ldots , u_{n} \in P$ and where we use an explicit notation for
$\Delta_R$ on $P$..

Also automatically, the inclusion $j:B\hookrightarrow P$ extends to an
inclusion $j: \Omega B
\hookrightarrow \Omega P$. We will be especially interested in $\Gamma_{P}$ the
space of 1-forms on $P$. The natural
$P$-sub-bimodule here is
\eqn{hor}{\Gamma_{hor} =  Pj(\Gamma_{B})P\subseteq \Gamma_P}
where $\Gamma_B$ is the space of 1-forms on $B$. Here we think of
$\Gamma_{hor}$ as analogous to the space of horizontal forms coming in the
classical case by pull-back from the base. We say that a one-form $\alpha \in
\Gamma_{P}$ is {\em horizontal}
if $\alpha \in \Gamma_{hor}$.
% and {\em vertical} if $\alpha \in \Gamma_{ver}$. If there exists a
% connection in $P$ then any one-form $\alpha \in \Gamma_{P}$ can be
% uniquely written as a sum of a horizontal and a vertical forms.
Obviously any $\beta\in \Gamma_B$ is by
definition horizontal
when viewed in $\Gamma_{hor}$ via the canonical inclusion $j$.

Finally, we need the notion of a map $\widetilde{\  }$ generating the
fundamental vector fields for our coaction $\Delta_R$. This appears in our dual
formulation as a left $P$-module map
\eqn{tilde}{ \widetilde{\  }=(\cdot\otimes id) \circ (id\otimes
\Delta_R)|_{P^2}:
\Gamma_P \rightarrow
P\otimes A.}
Recall that by definition in the universal case $\Gamma_P$ is the set
$P^2\subset P\otimes P$ where $P^2$ is the kernel of the product map. In
explicit terms we have
\eqn{tilde.udv}{\widetilde{\  }(udv)=\sum u v\bo\tens v\bt-uv\tens 1.}
Because $A$ coacts on $P$ from the right, $A^*$ acts on $P$ from the left. The
action of
$\xi\in A^*$  is given by evaluation against the output of the coaction. Hence
the left
$P$-module map   $\widetilde\xi=(\id\tens\xi)\circ\widetilde{\  }:\Gamma_P\to
P$
should be thought of as the `fundamental vector field' generated by
the `infinitesimal' element $\xi-1\eps(\xi)$. Compare
(\ref{classical.tilde}). It is also easy to see from these definitions that
\eqn{horizsubkernerl}{\ker\widetilde{\  }\supseteq \Gamma_{hor}.}
This is because
\begin{eqnarray*}
\widetilde{\  }(u(dj(b))v) & = & \widetilde{\  }(ud(j(b)v))-\widetilde{\
}(uj(b)dv)\\
& = & \sum
uj(b)\bo v\bo\tens j(b)\bt v\bt- \sum uj(b)v\bo\tens v\bt=0
\end{eqnarray*}
where the first equality  uses the Leibniz rule in $\Gamma_P$ and
second that $P$ is a comodule algebra.

We  are now ready to present the construction of quantum fibre bundles
and connections on them.
\begin{df}\label{principal.bundle}
We say that $P = P(B,A)$ is a {\em quantum principal
bundle} with universal differential calculus, structure quantum group $A$ and
base $B$ if:
\begin{enumerate}
\item $A$ is a Hopf algebra.
\item $(P, \Delta_{R})$ is a right $A$-comodule algebra.
\item $B = P^{A} = \{ u \in P : \Delta_{R} u = u \otimes 1 \}$.
\item $(\cdot\tens\id)(\id\tens\Delta_R):P\tens P\to P\tens A$ is a surjection
(freeness condition).
\item $\ker\widetilde{\  }=\Gamma_{hor}$ (exactness condition for the
differential envelope).
\end{enumerate}
% We suppose that any desired $(\Omega (P),d)$ is covariant in the sense of
% (\ref{covariance}).
\end{df}

The last condition here needs some explanation. In the classical case
smoothness and dimension considerations combine with freeness of the action to
ensure that the quotient is a manifold and the fiber through a point $u$ is a
copy of our Lie group $G$. At the differential level the Lie algebra $g$ of $G$
is included in the vertical part of $T_uP$ by the map
$\widetilde{\  }$ that generate fundamental vector fields. Dimension
arguments then imply that this map is an isomorphism of $g$ with the vertical
part of each $T_uP$. In our algebraic formulation we need to impose some kind
of condition to replace this complex of ideas arising from the smoothness and
dimension considerations. The one stated in the definition appears the most
convenient for our formulation below. Other approaches are surely possible
also. Roughly speaking in place of dimension arguments we suppose directly that
the image of the fundamental vector fields through each point span all the
vertical vectors through the point. Put another way in terms of forms, we
suppose that the horizontal forms span all of the anihilator the left-invariant
vector fields. In dual form this leads to the condition 5 in the definition.
We call it exactness because it states that the image of $j$ fills out the
kernel of $\widetilde{\ }$. It is stated here for the case of the universal
differential envelope on $P$.

We note also that this exactness condition is a kind of differential version of
the idea of a Galois extension in algebra. Given
conditions 1.-3 as above it is easy to see that the canonical map
$(\cdot\tens\id)(\id\tens\Delta_R):P\tens P\to P\tens A$ descends to a map
$P\tens_B P\to
P\tens A$ and $B\subset P$ is called a Galois extension if the map at this
level is an isomorphism, see e.g.\cite{Sch:pri}. Surjectivity corresponds to
our freeness condition and
injectivity is sufficient to prove exactness in our sense. This is because
$\widetilde{\ }$ is the canonical map restricted to $P^2\subset P\tens P$.
Hence an element of its kernel is also in the kernel of the canonical map and
hence, in the Galois case, in the kernel of the projection $P\tens P\to
P\tens_B P$. But the kernel of the restriction of this map to $P^2$ can be
identified with $Pj(B^2)P=\Gamma_{hor}$. On the other hand our geometrical
condition is weaker and moreover, in a form that is suitable for generalisation
later to non-universal differential calculi.

\begin{ex}\label{trivial.bundle}
Let $A$ be a Hopf algebra and $P$ an $A$-comodule algebra with invariant
subalgebra $B$. Suppose that there exists a convolution invertible  map $\Phi :
A
\hookrightarrow P$ such that
\begin{equation}
\Delta_{R} \circ \Phi = ( \Phi \otimes id) \circ \Delta ,
\hspace{1cm} \Phi (1_{A}) = 1_{P}
\label{trivialization}
\end{equation}
(so $\Phi$ is an intertwiner for the right coaction). Then $P$ is a quantum
principal bundle. We call $P(B,A,\Phi)$ a trivial bundle with {\em
trivialization} $\Phi$.
\label{def.principal.trivial}
\end{ex}
\proof An elementary fact in the situation of the example is that the
map
\eqn{trivial.iso}{ B \otimes A \rightarrow P,\qquad b \otimes a \mapsto
j(b)\Phi(a)}
is an isomorphism of linear spaces. Explicitly the inverse is given by
\[ u\mapsto \sum u\bo\Phi^{-1}(u\bt\o)\tens u\bt\t.\]
Using that $\Phi$ is an intertwiner and the properties of comodule algebras
etc as in Section~3 we observe that
\eqn{coaction.phiinv}{\Delta_{R} \Phi ^{-1} (a) = \Phi ^{-1} (a_{(2)}) \otimes
Sa_{(1)}}
after which it is clear that the image of our inverse map lies in $B\tens A$.
It is then
easy to verify that it provides the necessary inversion.

{}From this it follows that the freeness and exactness conditions 4. and 5. in
Definition~\ref{principal.bundle} are automatically satisfied in this case.
For the first condition assume that $\sum u_k\otimes a^k \in P\otimes A$.
Define
an element
$\rho\in\Gamma_P$ by
\[\rho = \sum
u_k\Phi^{-1}(a^k\o)\tens\Phi(a^k\t) .\]
Then
\begin{eqnarray*}
(\cdot\tens\id)(\id\tens\Delta_R)(\rho) & = & \sum
u_k\Phi^{-1}(a^k\o)\Phi(a^k\t)\otimes
a^k_{(3)} = \sum u_k\otimes a^k.
\end{eqnarray*}
The last equality follows from the intertwiner property (\ref{trivialization}).
Hence the coaction is free.

For the exactness condition we have to show that $\ker\widetilde{\  }=Pd
j(B)P$ where $d$ is the universal differential as recalled in Section~2 and
we work with $\Gamma_P$ as the subspace $P^2$ of $P\tens P$. Now any element
$\rho\in\ker\widetilde{\  }$ can be written as $\rho=\sum_i u_i dv_i$ for
$u_i,v_i\in P$. But since $\Phi$
establishes an isomorphism between $P$ and $B\tens A$ we can write each
$v_i=\sum_k j(b_i^k)\Phi(a_i^k)$. Applying $\widetilde{\ }$ to $\rho$ in this
form we deduce that
\[ 0=\widetilde\rho=\sum_{i,k} u_ij(b_i^k)\Phi(a_i^k\o)\tens
a_i^k\t-u_ij(b_i^k)\Phi(a_i^k)\tens 1\]
where we used that $\Phi$ is an intertwiner. Applying the map
$(\Phi^{-1}\tens\Phi)\circ\Delta$ to the second factor we obtain
\[  0=\sum_{i,k} u_ij(b_i^k)\Phi(a_i^k\o)\tens \Phi^{-1}(a_i^k\t)\tens
\Phi(a_i^k\th)-u_ij(b_i^k)\Phi(a_i^k)\tens 1\tens 1.\]
Finally we multiply the first two factors to conclude that
\[ 0=\sum_{i,k} u_ij(b_i^k)\tens \Phi(a_i)-u_ij(b_i^k)\Phi(a_i^k)\tens
1=\sum_{i,k} u_i j(b_i^k)d\Phi(a_i^k).\]
Hence using the Leibniz rule we have $\rho=\sum_{i,k} u_i
d(j(b_i^k)\Phi(a_i^k))=
\sum_{i,k} u_i (d(j(b_i^k)) \Phi(a_i^k)+ u_i j(b_i^k) d\Phi(a_i^k)=\sum_{i,k}
u_i (dj(b_i^k))\Phi(a_i^k)$ and hence manifestly lies in $Pdj(B)P$ as
required. \endproof

Next in our dual formulation we define a {\em connection} $\Pi$ on a quantum
principal bundle $P$ as an
assignment of a left $P$-submodule $\Gamma_{ver}\subseteq\Gamma_{P}$
such that:
\begin{enumerate}
\item $\Gamma_{P} = \Gamma_{hor} \oplus \Gamma_{ver}$,
\item  projection $\Pi : \Gamma_{P} \rightarrow \Gamma_{ver}$ is right
invariant  i.e.
\begin{equation}
\Delta_{R} \Pi = ( \Pi \otimes id ) \Delta_{R}.
\label{inv.conn}
\end{equation}
\end{enumerate}

An element $\alpha \in \Gamma_{ver}$ is called  a {\em vertical
form}.  If there exists a
connection in $P$ then any one-form $\alpha \in \Gamma_{P}$ can be
uniquely written as a sum of a horizontal and a vertical forms.

We show now that every connection has a connection form. Notice first that the
space $P\otimes\ker\eps$ has a natural
left P-module structure. Moreover there is a natural right coaction of
$A$ on $P\otimes\ker\eps$ built up as follows. $A$ coacts on $P$ by
$\Delta_R$ and $A$ coacts on itself  by the right adjoint coaction
\eqn{def.Ad}{Ad_{R} : A
\rightarrow A\otimes A,\qquad Ad_R(a)=\sum a_{(2)} \otimes
(Sa_{(1)})a_{(3)}.}
It is easy to see that this restricts to a coaction $Ad_R$
on $\ker\eps$ also. Hence we may define
the right coaction
$\Delta_R:P\otimes\ker\eps\rightarrow P\otimes\ker\eps\otimes A$ by
\begin{equation}
\Delta_R(u\otimes a)= \sum u^{(\overline 1)}\otimes a\t\otimes
u^{(\overline 2)}(Sa\o)a_{(3)}.
\end{equation}

We will need the following
\begin{lemma}\label{tilde.intertwiner}
The map $\widetilde{\  }$ intertwines right coactions on $\Gamma_P$
and $P\tens\ker\eps$,
\begin{equation}
\Delta_R\widetilde{\  } = (\widetilde{\  } \otimes id)\Delta_R
\end{equation}
\end{lemma}
\proof It is immediate from the form (\ref{tilde.udv}) that the image of
$\widetilde{\ }$ lies in $P\tens \ker\eps$. For any $\sum u_k\otimes v_k \in
\Gamma_P$ we have
\begin{eqnarray}
\Delta_R\widetilde{\  }(\sum u_k\otimes u_k) & = & \Delta_R (\sum
u_kv_k\bo\otimes v_k\bt) \nonumber \\
& = & \sum u_k\bo v_k\bo\otimes
{v_k\bt}_{(3)}\otimes u_k\bt{v_k\bt}\o (S{v_k\bt}\t){v_k\bt}_{(4)}
\nonumber \\
& = & \sum u_k\bo v_k\bo\otimes {v_k\bt}_{(1)}\otimes u_k\bt
{v_k\bt}_{(2)}. \nonumber
\end{eqnarray}On the other hand
\begin{eqnarray*}
(\widetilde{\  }\otimes id)\Delta_R (\sum u_k\otimes v_k) & = &
(\widetilde{\  }\otimes id) (\sum u_k\bo\otimes v_k\bo\otimes u_k\bt v_k\bt
)\\
& = & \sum u_k\bo v_k\bo\otimes {v_k\bt}\o\otimes u_k\bt {v_k\bt}\t .
\end{eqnarray*}
as required. \endproof

The freeness and exactness conditions imply that the
following sequence
\begin{equation}
0\rightarrow\Gamma_{hor}\buildrel{j}\over{\rightarrow}\Gamma_P
\widetilde{\rightarrow}P\otimes\ker\eps\rightarrow 0
\label{sequence}
\end{equation}
is exact. The existence of the connection $\Pi$ in $P$ is  now
equivalent to the existence of the map $\sigma
:P\otimes\ker\eps\rightarrow\Gamma_P$ splitting  the sequence
(\ref{sequence}), i.e. $\widetilde{\  }\circ\sigma = id$. Due to the fact
that $\Pi$ is a right-invariant left $P$-module map, the map $\sigma$
has to be a right-invariant left $P$-module map. The projection $\Pi$ is
recovered as
\begin{equation}
\Pi = \sigma\circ\widetilde{\ } .
\end{equation}

Now we define a map $\omega : A\rightarrow\Gamma_P$ by
\begin{equation}
\omega(a) = \sigma (1\otimes (a-\eps(a))).
\label{def.omega.sigma}
\end{equation}
We call this map the {\em connection form} of the connection $\Pi$.

\begin{prop}\label{prop.connection} Let $P$ be a quantum principal bundle and
$\Pi$ a connection on it. Then the connection form  $\omega :
A\rightarrow\Gamma_P$ has the following properties.
\begin{enumerate}
\item $\omega(1)=0$
\item $\widetilde{\  }\omega(a)=1\tens a-1\tens 1\eps(a)$ for all $a\in A$
\item $\Delta_R\circ\omega=(\omega\tens\id)\circ Ad_R$
\end{enumerate}
\noindent where $Ad_R$ is the right adjoint coaction. Conversely if
$\omega$ is any linear map
$\omega:A\rightarrow\Gamma_P$ obeying conditions 1.-3. then
there is a unique connection $\Pi$,
\eqn{connection.1.form}{ \Pi= \cdot \circ (id \otimes \omega) \circ
\widetilde{\ }}
such that  $\omega$ is its connection 1-form.
\end{prop}
\proof Given $\Pi$ we define $\omega(a)=\sigma(1\tens (a-\eps(a))$ as explained
above. Then properties 1. and 2. follow immediately from the
definition of $\omega$.

Next we have to show that $\omega$ is $Ad_R$-covariant. We have
\begin{eqnarray*}
\Delta_R(\omega (a)) & = & \Delta_R\sigma(1\otimes (a-\eps(a))) \\
& = & (\sigma\otimes id)\Delta_R (1\otimes (a-\eps(a))) \\
& = & \sum (\sigma\otimes id) (1\otimes (a\t-\eps(a\t))\otimes (Sa\o
)a_{(3)}) \\
& = & \sum \omega (a\t)\otimes (Sa\o)a_{(3)}.
\end{eqnarray*}
{}From this it follows at once that $\omega$ obeys the equivariance condition
3.

In the converse direction suppose that we are given a map $\omega$ obeying
conditions 1.-3. and define $\sigma:P\tens\ker\eps\to\Gamma_P$ by
$\sigma(u\tens a)=u\omega(a)$ for $u\in P$ and $a\in\ker\eps$.
Then $\widetilde{\  }\circ\sigma(u\tens a)=u\widetilde{\  }(\omega(a))=u\tens
a$ by the first condition on $\omega$. Hence $\widetilde{\ }\circ\sigma=\id$
and $\Pi=\sigma\circ\widetilde{\ }$ is a splitting
$\Gamma_P=\Gamma_{hor}\oplus\Gamma_{ver}$ as required for a connection.
Explicitly,
\begin{eqnarray*}
\Pi(udv)=\sigma\circ\widetilde{\  } (udv) & = & \sum\sigma(uv\bo\otimes
(v\bt-\eps(v\bt))) \\
& = & \sum uv\bo\omega (v\bt -\eps(v\bt)) =  \cdot \circ (id \otimes
\omega) \circ \widetilde{\ }(udv).
\end{eqnarray*}
which is the form stated. Note that one can easily see directly that $\Pi$
defined via
(\ref{connection.1.form}) is a projection and $\ker\Pi\supseteq\Gamma_{hor}$,
but its description in terms of splitting as here is made possible by
the exactness condition as explained above.

One can also see that $\sigma$ as defined is equivariant if $\omega$ obeys
condition~3. From this it follows that $\Pi$ is also. For a direct proof, if
$\omega$ intertwines $\Delta_R$ and $Ad_R$ then
\align{\Delta_R\Pi(udv)&&=\sum u\bo v\bo\bo\omega(v\bt)\bo \tens
u\bt v\bo\bt\omega(v\bt)\bt \\
&&=\sum v\bo\omega(u\bo v\bt\t)\bo \tens u\bt v\bt\o\omega(v\bt\t)\bt \\
&&=\sum u\bo v\bo\omega(v\bt\t\t)\tens u\bt v\bt\o (S v\bt\t\o)v\bt\t\th\\
&&=\sum u\bo v\bo\omega(v\bt\o)u\bo\tens u\bt v\bt\t \\
&&=\sum u\bo
v\bo\bo\omega(v\bo\bt)\tens u\bt v\bt \\
&&=\sum \Pi(u\bo(dv\bo))\tens u\bt v\bt
=(\Pi\tens\id)\circ\Delta_R(udv)}
as required. We use that $\Delta_R$ is a comodule algebra, the
intertwiner property of $\omega$, the antipode axioms and finally that
$\Delta_R$ is a comodule algebra again. \endproof

The condition~2. in the
proposition is analogous to the classical condition that $\omega$
behaves like  the Maurer-Cartan form when evaluated on fundamental
vector fields.  The condition 2. is analogous to its usual
$Ad$-equivariance property.  The proposition tells us how we can
manufacture connections  from connection one-forms.

\begin{ex}
Let $P(B,A,\Phi)$ be  the trivial quantum principal bundle in
Example~\ref{trivial.bundle}.
There is a natural connection
$\Pi_{triv}$ given by the connection  1-form $\omega_{triv}(a) = \sum \Phi^{-1}
(a_{(1)})d \Phi (a_{(2)})$. For this connection we have
\[ \Gamma_{ver}=Pd\Phi(A)\equiv \{ud\Phi(a):u\in P\ a\in A\}\]
and the splitting $\Gamma_P=\Gamma_{hor}\oplus \Gamma_{ver}$ is
according to the  Leibniz rule in $\Gamma_P$,
\[ ud(j(b)\Phi(a))=u(dj(b))\Phi(a)+uj(b)d\Phi(a) \in P j(\Gamma_{B})P
\oplus P d\Phi(A) \]
\label{trivial.connection}
\end{ex}
\proof Firstly we compute
\begin{eqnarray*}
\widetilde{\  }\omega_{triv}(a) & = & \sum \widetilde{\
}(\Phi^{-1}(a\o)d\Phi(a\t)) =
\sum \Phi^{-1}(a\o)\Phi(a\t)\bo\tens
\Phi(a\t)\bt-\eps(a)\tens 1\\
& = & 1\tens a-\eps(a)\tens 1
\end{eqnarray*}
using right-invariance of the co-ordinate chart $\Phi$.

Secondly we show that the map $\omega_{triv} (a)$ is an
intertwiner between the adjoint coaction and the right coaction
$\Delta_{R}$ of
$A$ on $P$. Using (\ref{coaction.phiinv}) we have
\begin{eqnarray*}
\Delta_{R} \omega_{triv} (a) & = & \Delta_{R} \sum (\Phi ^{-1}(a_{(1)})d\Phi
(a_{(2)}))  =  \sum \Phi^{-1}(a_{(2)})d\Phi(a_{(3)}) \otimes
(Sa_{(1)})a_{(4)} \\
& = & (\omega_{triv} \otimes id) Ad_{R} (a).
\end{eqnarray*}
as required. Obviously $\omega_{triv} (1) = 0$.

Hence by
Proposition \ref{prop.connection} we conclude that we have a connection
$\Pi_{triv}$ with
$\omega_{triv}$ as its connection form. To compute $\Gamma_{ver}$ we have
\begin{eqnarray*}
\Pi_{triv}(ud\Phi(a)) & = &\sum u\Phi(a)\bo\omega_{triv}(\Phi(a)\bt)=\sum
u\Phi(a\o)\omega_{triv}(a\t) \\
& = & \sum u\Phi(a\o)\Phi^{-1}(a\t)d\Phi(a\th) = ud\Phi(a)
\end{eqnarray*}
so that $Pd\Phi(A)\subseteq \image \Pi_{triv}$. Next, applying
$\Pi_{triv}$ to a general  element of $\Gamma_P$ we have
\begin{eqnarray*}
 \Pi_{triv}(\sum ud(j(b_i)\Phi(a_i))) & = & \Pi_{triv}(\sum
u(dj(b_i))\Phi(a_i)+\sum uj(b_i)d\Phi(a_i)) \\
& = & \Pi_{triv}(\sum
uj(b_i)d\Phi(a_i)) = \sum uj(b_i)d\Phi(a_i).
\end{eqnarray*}
The element $\sum u(dj(b_i))\Phi(a_i)$ here is
manifestly horizontal  and hence annihilated by $\Pi_{triv}$. This
shows that $Pd\Phi(A)=\image \Pi_{triv}$.
\endproof

\vspace{12pt}

Thus every trivial bundle has a canonical trivial connection. More generally we
have the following construction that gives the relationship between a
connection
1-form $\omega$ as above and a connection 1-form $\beta$ as defined in
the previous section.

\begin{prop}\label{beta.prop}
Let $\beta : A \rightarrow \Gamma_{B}$ be a linear map such that
$\beta (1)
= 0$. Then the map
\begin{equation}
\omega (a) = \sum \Phi ^{-1}(a_{(1)}) j(\beta (a_{(2)})) \Phi
(a_{(3)}) + \sum \Phi ^{-1}(a_{(1)}) d \Phi (a_{(2)})
\label{trivial.connection.beta}
\end{equation}
is a connection 1-form in the trivial principal bundle $P(B,A,\Phi)$ with
trivialization $\Phi$.
\end{prop}
\proof Note that the last part of (\ref{trivial.connection.beta})
coincides with the connection 1-form $\omega_{triv}$ defined in
Example~\ref{trivial.connection}. We have now
\begin{eqnarray*}
\Delta_{R} \omega (a) & = & \sum (\Phi ^{-1}(a_{(2)}) j(\beta
(a_{(3)})) \Phi (a_{(4)}) \otimes (S a_{(1)}) a_{(5)} + \Phi ^{-1}
(a_{(2)}) d\Phi(a_{(3)}) \otimes (Sa_{(1)}) a_{(4)}) \\
& = & ((\Phi^{-1} * (j \circ \beta) * \Phi + \omega_{triv} )\otimes 1 )
Ad_{R} (a).
\end{eqnarray*}
Hence $\omega $ is an intertwiner between $\Delta_{R}$ and
$Ad_{R}$. Applying the map $\widetilde{\ }$ to $\omega$ we see that the first
part of the sum (\ref{trivial.connection.beta}) is annihilated
(because it is horizontal). From
Example~\ref{trivial.connection}
we know that $\widetilde{\ } \omega_{triv} (a) = 1 \otimes a - \epsilon (a)
\otimes 1$
for any $a \in A$, hence the same is true for $\omega$. Hence by
Proposition~\ref{prop.connection} we can define connection $\Pi$.
\endproof

We note that in the case of a
trivial bundle  with  connection and
connection form $\omega$  as in the last proposition, one still has $
P\Phi(\Gamma_A)\isom \Gamma_{ver}$.
The isomorphism   means that every form in $\Gamma_A$ can be lifted to
a form in  $\Gamma_P$. The explicit formula is
\[ ud\Phi(a)\mapsto \Pi(u d\Phi(a))=\sum u\Phi(a\o)\omega(a\t).\]
This follows from the same techniques as in the proof above.

These connections also provide covariant derivatives on horizontal
pseudotensorial forms on $P$ defined as the differential followed by horizontal
projection. Moreover, these can also be understood as sections of associated
vector bundles etc. as in the classical theory. Details are given in the
Appendix A and justify further our present formalism.

Next we come to the important notion of gauge transformation of principal
bundles. Let $P(B,A,\Phi)$ be a trivial quantum principal bundle with
trivialization
$\Phi : A \rightarrow P$ and let $\gamma :A \rightarrow B$ be a
convolution invertible linear map such that $\gamma (1) = 1$.  We say
that the map
\begin{equation}
\Phi ^{\gamma} = \sum j(\gamma (a_{(1)}))\Phi (a_{(2)}) = ((j \circ
\gamma) * \Phi) (a)
\label{gauge.trans.1}
\end{equation}
is a {\em gauge transformation} of $\Phi$.

\begin{prop} \label{change.trivialization}
If $P(B,A,\Phi)$ is the trivial quantum principal bundle as in
Example~\ref{trivial.bundle} with trivialization $\Phi$, then $P(B,A,\Phi
^{\gamma})$  is also a trivial quantum principal bundle with trivialization
$\Phi^\gamma$ defined by (\ref{gauge.trans.1}).
\end{prop}
\proof Note that since $\gamma$ is a convolution invertible map,
$\Phi^{\gamma}$ is also convolution invertible. Moreover, $\Phi^\gamma
(1) = 1$.  We need only to check that $\Phi ^{\gamma}$ is an intertwiner. We
have
\begin{eqnarray*}
\Delta_{R} \Phi ^{\gamma}(a) & = & \sum \Delta_{R} (j(\gamma (a_{(1)}))
\Phi (a_{(2)}) = \sum j(\gamma(a_{(1)})) \Phi (a_{(2)}) \otimes
a_{(3)} = (\Phi ^{\gamma} \otimes id)\Delta (a).
\end{eqnarray*}
In the second equality we have used the interwiner property of $\Phi$ and
the fact that $j \circ \gamma(a)$ is in the invariant part of $P$.
Hence $\Phi ^{\gamma}$ is a trivialization of $P$. \endproof

The proposition gives the interpretation of a gauge transformation as a
change of local coordinates in $P$. Next we see that a gauge
transformation induces a corresponding transformation of a
connection 1-form $\beta$ on our trivial quantum vector bundle. We have the
following:
\begin{prop}\label{gauge.transform.beta}
Let $P$, $\beta$ and $\omega$ be as in Proposition~\ref{beta.prop}.
Let $\gamma : A \rightarrow B$ be a gauge
transformation. The transformation $\beta \mapsto \beta ^\gamma$,
\begin{equation}
\beta ^{\gamma} = \gamma ^{-1} * \beta * \gamma + \gamma ^{-1} *
d\gamma
\label{beta.transform.1}
\end{equation}
for fixed $\Phi$ induces a transformation $\omega \mapsto
\omega^\gamma$ which can be understood as a gauge transform
$\Phi\to\Phi^\gamma$ for fixed $\beta$,
\[ \omega^\gamma = (\Phi^{\gamma})^{-1} * j(\beta) * \Phi ^\gamma + (\Phi
^{\gamma})^{ -1}* d\Phi^\gamma. \]

Conversely, for fixed $\omega$ the change of trivialization  $\Phi$  by a gauge
transformation $\gamma$ induces the transformation $\beta \mapsto
\beta^{\gamma^{-1}}$ where
\begin{equation}
\beta^{\gamma^{-1}}= \gamma * \beta * \gamma ^{-1} + \gamma * d\gamma ^{-1}
\end{equation}
\end{prop}
{\bf Proof} This  follows by direct computation. The first statement is
\begin{eqnarray*}
\omega ^\gamma & = &\Phi^{-1} * j(\beta ^\gamma) * \Phi  + \Phi ^{-1}*
d \Phi \\
& = & \Phi ^{-1} * j(\gamma^{-1}) * j(\beta) * j(\gamma) *\Phi + \Phi
^{-1} *j(\gamma^{-1})*(dj(\gamma ))* \Phi  + \Phi^{-1}* d \Phi \\
& = & (\Phi ^{\gamma})^{ -1} * j(\beta) * \Phi ^\gamma + \Phi^{-1} *
j(\gamma ^{-1}) * d(j(\gamma) * \Phi ) - \Phi ^{-1} * d\Phi + \Phi
^{-1} *d\Phi \\
& = &(\Phi ^{\gamma})^{ -1} * j(\beta) * \Phi ^\gamma + (\Phi ^{\gamma})^{ -1}
*d\Phi ^\gamma.
\end{eqnarray*}
Note that thanks to Proposition~\ref{beta.prop}, $\omega^\gamma$
is a
connection 1-form.

To prove the converse we have
\begin{eqnarray*}
\omega & = & (\Phi ^{\gamma})^{ -1} * j(\beta ')* \Phi ^\gamma +
(\Phi^{\gamma})^{ -1} * d\Phi^\gamma \\
& = &\Phi^{-1} * j(\gamma^{-1} * \beta ' * \gamma)* \Phi + \Phi^{-1} *
j(\gamma ^{-1} )* d(j(\gamma) * \Phi) \\
& = & \Phi ^{-1} * j(\gamma ^{-1} * \beta '* \gamma) * \Phi + \Phi^{-1}*
(j(\gamma ^{-1})* dj(\gamma)) *\Phi+ \Phi ^{-1} * d\Phi.
\end{eqnarray*}
Comparing with  Proposition~\ref{beta.prop} this means that $\beta'$
necessarily obeys
\[ \Phi^{-1} * j(\gamma^{-1}* \beta '* \gamma )* \Phi + \Phi^{-1}
*j(\gamma^{-1}* d\gamma)*\Phi = \Phi^{-1} * j(\beta) * \Phi \]
which is equivalent to $\beta' = \gamma * \beta * \gamma^{-1} + \gamma * d
\gamma ^{-1}$ by conjugating by $\Phi$ in the convolution algebra. Thus the
effect of a gauge-transformation does not take us out of the class of
connections of the form of  Proposition~\ref{beta.prop} and the required
transformation of $\beta$ is uniquely determined.
\endproof

In the same way the gauge transformation of quantum associated vector bundles
and their sections are induced by a change of trivialization $\Phi$. These
details are included for completeness in Appendix A and tie up the present
formulation precisely with the elementary local picture in Section~3.

Finally, now that we understand properly the notion of trivial bundles and
their
gauge transformation
properties we are in a position to introduce
the notion of a {\em locally trivial quantum bundle} as a collection
of trivial bundles pasted
together via gauge transformations. This is exactly in analogy with
the usual definition
of local trivializations of principal bundles except, of course, that
we must work algebraically
as in sheaf theory, and that by gauge transforms we mean the convolution by
convolution-invertible maps as
in Proposition~\ref{change.trivialization}. Thus, the most naive formulation of
a
locally-trivial principle bundle consists of the
following data.

1. An index set $I=\{i,j,ij\cdots\}$ to be thought of as labeling the
members of an `open cover',  with analogous properties. There should
be a partial ordering (corresponding to inclusion) and a product
(corresponding to intersection) with $ij\le i,j$. Indexed by this,
we consider a collection of algebras $P_i$ with maps $P\to P_i$
and $P_i\to P_j$ for $i\ge j$ (the {\em restriction maps}) and
the equalizer
\[ P\to \prod P_i{\to\atop \to}\prod P_{ij}.\]
We mean here the usual picture in sheaf theory (see for example
\cite[Sec. 2.2]{BarWel:top}) so that if $u_i\in P_i$ are given such
that their restrictions to each $P_{ij}$ coincide then they are
themselves the restriction of some $u\in P$. The  algebras $P_i$ are
each $A$-comodule algebras (and the restriction maps are
intertwiners), and $B_i=P_i^A$ are such that $B\to \prod B_i{\to\atop
\to}\prod B_{ij}$.

2. There are trivializations $\Phi_i:A\to P_i$ making $B_i\subseteq
P_i$ trivial bundles.

3. There are  convolution-invertible maps $\gamma_{ij}:A\to B_{ij}$
such that \[ \sum \gamma_{ij}*\gamma_{jk}=\gamma_{ik},\quad
\Phi_i= \sum \gamma_{ij} * \Phi_j\]
where the maps are composed with the relevant restriction maps such
that the results are maps $A\to B_{ijk}$ and $A\to P_{ij}$ respectively.

This is the most naive definition based on the transformation
properties studied above. Note that in algebraic geometry, the ring
of functions on the open set consisting of the space minus a number of point is
achieved by inverting the points, i.e. by localization, and in this case the
corresponding restriction maps are inclusions.  While adequate to cover our
examples in Section~5, it should be noted that this it is not the only possible
formulation. Also, the index set could have
properties somewhat weaker than those of a classical open cover. It is expected
that a rather
bigger repertoire of non-commutative examples will be needed before the most
suitable direction for a complete formulation can be determined.

\subsection{The Case of General Quantum Differential Calculi}

The theory above has been developed for simplicity in the case of the universal
differential envelope on $P$. This made contact with the local picture of
connections defined by one-forms on the base and gauge transformations as in
Section~3.
Now we give the further refinements needed for the
non-universal case. We have to suppose differential structures on both $P$ and
$A$ and suitable compatibility conditions between them. This refinement is
needed to make contact with examples that truly deform the
usual commutative differential calculus, such as our monopole example of
Section~5.

We begin with a few words about the general theory of bicovariant differential
calculi
on quantum groups \cite{Wor:dif}. A bicovariant differential
calculus on a quantum group $A$ is a pair $(\Gamma_A ,d )$ such that
$\Gamma_A$ is a left and right $A$-comodule and $d$ is a comodule map,
i.e.
\[\Delta d = (d\otimes id)\Delta_R \quad \Delta d = (id \otimes
d)\Delta_L\]
where $\Delta_R$ and $\Delta_L$ are right and left coactions of $A$ on
$\Gamma_A$. If $\Gamma_A$ is a left (right) $A$-comodule only and $d$
is a comodule map then $(\Gamma_A,d)$ is called a left-covariant
(right-covariant) differential calculus on $A$. The universal differential
calculus on $A$ is an example of a
bicovariant differential calculus. The coactions of $A$ on $A^2$ are
given by
\[\Delta^U_R = (id\otimes id\otimes \cdot)\circ (id\otimes\tau\otimes
id)\circ (\Delta\otimes\Delta )\]
\[\Delta_L^U = (\cdot\otimes id\otimes id)\circ (id\otimes\tau\otimes
id)\circ (\Delta\otimes\Delta ).\]
Every bicovariant differential calculus on $A$ can be obtained from
the universal one by taking a quotient $\Gamma_A = A^2/N_A$ where
$N_A$ is a sub-bimodule of $A^2$ such that
\[\Delta^U_RN_A\subset N_A\otimes A \quad \Delta_L^UN_A\subset
A\otimes N_A.\]
Equally-well one can take a right ideal $M_A\subset\ker\epsilon$ such
that
\begin{equation}
Ad_RM_A\subset M_A\otimes A
\label{m.ad.invariance}
\end{equation}
and define $N_A = \kappa(A\otimes M_A)$ where the map $\kappa : A\tens
A \rightarrow A\tens A$, given by
\eqn{kappa.A}{\kappa (a\otimes a') =
\sum aSa'\o\otimes a'\t}
is a linear isomorphism. If the ideal does not obey
(\ref{m.ad.invariance}) then the resulting calculus is only left-covariant.
We will always assume that the differential structure on our quantum group $A$
is
bicovariant.

\note{Let $A$ be a quantum group. Assume that $(\Gamma_A,d)$ is at least
left-covariant. We say that $\rho\in\Gamma_A$
is left-invariant if $\Delta_L\rho = 1\otimes\rho$. If $\{\omega^i\}$,
$i\in I$ is a basis in the space of all left-invariant forms in
$\Gamma_A$  then every $\rho\in\Gamma_A$ can be written uniquely as
\begin{equation}
\rho = \sum_i \rho_i\omega^i
\end{equation}
where $\rho_i\in A$. In particular we can
write the differential $d$ in the form
\begin{equation}
da = \sum a_{(1)}\chi_i(a\t)\omega^i.
\end{equation}
where functionals $\chi_i$ generate the subspace $A_c^*\subset A^*$
given as
\[ A_c^* = \{\chi \in A^* :\forall a\in M_A\cup k\cdot 1 ,\;
\chi(a)=0 \}. \]}

Next we come to the differential structure on the
quantum principal bundle $P$. As explained in Section~2 it is sufficient to
give the first order
differential structure $\Gamma_P$ as a quotient of the universal one,
$\Gamma_P=P^2/N_P$ where $N_P$ is a sub-bimodule of $P^2$. We will always take
$\Gamma_P$ to be of this form.

For our first compatibility between these structures we need to suppose that
the right coaction of $A$ on $P$ for our quantum principal bundle extends to
right-covariance of $\Gamma_P$ in a natural way. Recall that this was automatic
in the universal case. A sufficient condition for the same formula
(\ref{covariance}) to project down to the non-universal case is clearly
\[ \Delta_R N_P\subset N_P\tens A.\]
Likewise we need that our map $\widetilde{\ }$ generating the fundamental
vector fields in (\ref{tilde}) projects down to the non-universal case. It is
easy to see that the relevant condition is
\[ \widetilde(N_P)\subset P\tens M_A.\]
In this case we have a well-defined map $\widetilde{\  }_{N_P}:
\Gamma_P\rightarrow P\otimes \ker\eps/M_A$ by
\begin{equation}
\widetilde{\  }_{N_P}(\rho) = (id\otimes\pi_A)\circ\widetilde{\  }(\rho_U).
\label{compatible}
\end{equation}
where $\pi_{N_P}:P^2\rightarrow \Gamma_P$ and $\pi_A:ker\eps\rightarrow\ker\eps
/M_A$ are the canonical epimorphisms and for $\rho\in\Gamma_P$ we can take any
representative
$\rho_U\in\pi_{N_P}^{-1}(\rho)$. Note that the image of $\widetilde{\ }$ in
(\ref{tilde}) is automatically in $\ker\eps$ and we are relying on this now to
project down to $\ker\eps/M_A$. This time the corresponding vector field
$\Gamma_P\to P$ is obtained by evaluation against an element of the dual of
this.

\note{Thanks to the second requirement of Definition~\ref{principal.bundle.gen}
the map $\widetilde{\  }_{N_P}$ is well-defined. Precisely, take $\rho_U
,\rho'_U\in \pi^{-1}_{N_P}(\rho)$. Then $\rho_U-\rho'_U \in N$ and
$\widetilde{\  }(\rho_U -\rho'_U)\in P\otimes M_A$. Therefore
$(id\otimes\pi_A)\circ\widetilde{\  }(\rho_U-\rho'_U)=0$, and definition of
$\widetilde{\  }_N(\rho)$ does not depend on the choice of a representative in
$\pi_N^{-1}(\rho)$. }

\begin{df}\label{principal.bundle.gen}
We say that $P=P(B,A,N_P,M_A)$ is a {\em quantum principal
bundle} with structure quantum group $A$ and base $B$ and quantum differential
calculi defined by $N_P,M_A$ if:
\begin{enumerate}
\item $A$ is a Hopf algebra.
\item $(P, \Delta_{R})$ is a right $A$-comodule algebra.
\item $B = P^{A} = \{ u \in P : \Delta_{R} u = u \otimes 1 \}$.
\item $(\cdot\tens\id)(\id\tens\Delta_R):P\tens P\to P\tens A$ is a surjection
(freeness condition).
\item $\Delta_R N_P \subset N_P\otimes A$ (right covariance of differential
structure).
\item $\widetilde{\  }(N_P)\subset P\otimes M_A$ (fundamental vector fields
compatibility condition)
\item $\ker\widetilde{\  }_{N_P}=\Gamma_{hor}$ (exactness condition).
\end{enumerate}
%We suppose that any desired $(\Omega (P),d)$ is covariant in the sense of
% (\ref{covariance}).
\end{df}

Now we can define the
notions of horizontal
1-forms, connections and connection 1-forms precisely as in the universal
case. Thus a connection is an equivaraint splitting of $\Gamma_P$. This time
the
freeness condition ensures in particular that
\[ Im\widetilde{\  }_{N_P} = P\otimes\ker\epsilon/M_A.\]
Observe next that $Ad_R\ker\eps\subset\ker\eps\otimes A$. Since $M_A$
is $Ad_R$-invariant (equation (\ref{m.ad.invariance})) we have a
right-adjoint coaction of $A$ on
$\ker\eps /M_A$ by
\begin{equation}
Ad_R(\pi_A(a)) = \sum \pi_A(a\t)\otimes(Sa\o)a_{(3)}
\end{equation}
where $a\in\ker\eps$. Using the same methods as in
Lemma~\ref{tilde.intertwiner} we prove that $\widetilde{\  }_{N_P}$ is an
intertwiner. Finally, we have the exact sequence
\begin{equation}
0\rightarrow\Gamma_{hor}\buildrel{j}\over{\rightarrow}\Gamma_P
\buildrel{\widetilde{\  }_{N_P}}\over{\rightarrow}P\otimes\ker\eps/M_A
\rightarrow 0
\label{exact.sequence.n}
\end{equation}
of left $P$-module maps. This sequence splits whenever there is a
connection on $\Gamma_P$. If we denote the splitting (section) by
$\sigma_{N_P}$, then we can define a connection 1-form by
\begin{equation}
\omega(a) = \sigma_{N_P}(1\otimes\pi_A(a-\eps(a))).
\end{equation}

Now we can generalise Proposition~\ref{prop.connection}.

\begin{prop}\label{prop.connection.gen} Let $P(B,A,N_P,M_A)$ be a quantum
principal bundle and $\Pi$ a connection on it. Then its connection
1-form $\omega :A\rightarrow \Gamma_P$ has the following properties.
\begin{enumerate}
\item $\omega(1)=0$ and $\omega(M_A)=0$
\item $\widetilde{\  }_{N_P}\omega(a)= 1\tens \pi_A(a - \eps(a))$ for all $a\in
A$
\item $\Delta_R\circ\omega=(\omega\tens\id)\circ Ad_R$
\end{enumerate}
\noindent where $Ad_R$ is the right adjoint coaction.  Conversely if
$\omega$ is any
linear map $\omega: A\rightarrow\Gamma_P$ obeying
conditions 1.-3. then there is a unique connection $\Pi$,
\eqn{connection.1.form.gen}{ \Pi= \cdot \circ (id \otimes
\omega)\circ\widetilde{\ }_{N_P}}
 such that $\omega$ is
its connection one-form.
\end{prop}
\proof
\note{ Note firstly that the meaning of condition~3. is that this is the
necessary and sufficient condition for $\omega$ to be viewed
equivalently as a map $\und\omega:\widetilde{\ }_{N_P}\to \Gamma_P$.
In these projected terms the formula for the connection in terms of
the connection 1-form is \[ \Pi= \cdot \circ (id \otimes
\und\omega)\circ\widetilde{\ }_{N_P}\] and condition 1. is
 $\widetilde{\ }_{N_P}\und\omega(a)=a$ for all $a\in M_A$. }
The proof for the most part follows just the same steps
as the proof of Proposition~\ref{prop.connection} but at the quotient level.
The map $\omega$ is extracted from the splitting defined by the connection
and is $Ad_R$-covariant because $\sigma_{N_P}$ is right invariant. In
the converse
direction suppose that we are given a map $\omega$ obeying conditions 1.
-3. Condition~1. means that $\omega$ projects to a map
$\ker\eps/M_A\to\Gamma_P$ so
that $\Pi$ as stated is well-defined. Likewise
$\sigma_{N_P}:P\tens\ker\eps/M_A\to\Gamma_P$
is well-defined by
$\sigma_{N_P}(u\tens a)=u\omega(a)$ for $u\in P$ and $a\in\ker\eps/M_A$.
Then $\widetilde{\  }_{N_P}\circ\sigma_{N_P}(u\tens a)=u\widetilde{\
}_{N_P}(\omega(a))=u\tens a$ by the second condition on $\omega$. The
remaining steps are likewise similar.
\note{Note that it us convenient to work with the map $\und\omega$ but one
can also work with the map $\omega:\ker\eps/M_A\to\Gamma_P$}
\endproof

\begin{ex}\label{trivial.bundle.gen}
Let $P(B,A,\Phi)$ be as in Example~\ref{trivial.bundle}. If in addition the
differential structures are such that $\Delta_R N_P \subset N_P\otimes A$ and
\[ \widetilde{\  } (N_P)=P\otimes M_A\]
then the remaining conditions in Definition~\ref{principal.bundle.gen} are
automatically satisfied. We call this the trivial principal bundle with
trivialization $\Phi$ and general quantum differential calculus.
\end{ex}
\proof The freeness condition is already proven in Example~\ref{trivial.bundle}
and applies just as well here. For the exactness condition we also know that
$\ker\widetilde{\ }=P(d_U B)P$ (exactness in the universal calculus) from the
proof there. Take $\rho\in \ker\widetilde{\ }_{N_P}$ and choose a
representative $\rho_U\in\pi_{N_P}^{-1}(\rho)$. From the definition of
$\widetilde{\ }_{N_P}$ this means that $\widetilde{\ }(\rho_U)\in P\tens M_A$.
By our stronger version of the fundamental vector fields compatibility
condition as stated,
we know that there exists $\rho'_U\in N_P$ with $\widetilde{\
}\rho'_U=\widetilde{\ }\rho_U$. Hence by the exactness condition in the
universal differential envelope, we conclude  $\rho_U-\rho'_U\in P(d_UB)P$.
Since  $\rho=\pi_{N_P}(\rho_U-\rho'_U)$ we see that $\rho\in
P(dB)P=\Gamma_{hor}$ as required. \endproof

The slightly stronger form of the fundamental vector fields compatibility
condition (equality
rather than merely an inclusion) certainly
holds for usual trivial bundles with commutative differential calculi, as well
as for the trivial bundles (and also some non-trivial ones) constructed for
general differential calculi in the next section, i.e. in all known examples.
Hence it is natural to require it here for trivial bundles with general
calculi. Clearly other formulations are also possible. If we use this
formulation then we can also prove the existence of trivial connections on
trivial bundles. These can be constructed as follows. Let $\{e^i\in\ker\eps\}$
be such that $\{\pi_A(e^i)\}$ form a basis of $\ker\eps/M_A$ and for any $a\in
A$ write $\pi_A(a-\eps(a))=\sum_i c_i(a)\pi_A(e^i)$ say. Then
\[\omega(a)=\sum_i c_i(a)\Phi^{-1}(e^i\o)d\Phi(e^i\t)\]
is a connection with corresponding splitting according to the Leibniz rule (as
for the trivial connection in Example~\ref{trivial.connection})
at least on the elements corresponding to the basis,
\[\Pi \left(uj(b)\Phi(e^i)\right)=uj(b)d\Phi(e^i),\quad
(1-\Pi)\left(uj(b)\Phi(e^i)\right)=u(dj(b))\Phi(e^i)\]
We see here a significant complication caused by working with general
quantum-differential calculi: unless $\Phi$ is required to obey further
conditions the different choices of bases $\{e^i\}$ need not give the same
connection $\omega$. For example a sufficient condition for uniqueness of the
connection defined in this way is to assume that
\eqn{phi.nice}{\forall a\in M_A,\qquad \sum
\Phi^{-1}(a\o)\tens\Phi(a\t)\in N_P}
in which case all choices of basis give $\omega(a)=\sum
\Phi^{-1}(a\o)d\Phi(a\t)$. This condition is in turn implied in the commutative
case by the condition that $\Phi$ is an algebra map. On the other hand for a
quantum principal bundle we have already seen in Section~3 that one cannot
assume that $\Phi$ is an algebra map because this is not closed under
convolution, hence such a notion of trivial bundle could not be gauge
transformed. Likewise, the above slightly weaker condition
(\ref{phi.nice}) is not closed under
gauge transformation (i.e. if $\gamma$ and $\Phi$ obey it then $\Phi^\gamma$
need not).

This is also the reason that we limit ourselves in Section~3 and Appendix A to
the universal differential calculi. In fact, the general constructions in
Section~3 are self-contained and can be verified for any algebras and
differential calculi so long as we need only a local picture. For this picture
to come by association to a geometrical theory of principal bundles we have
to live with a certain amount of non-uniqueness or else impose further
conditions. Likewise, the notion of patching together trivial bundles as
outlined at the end of Section~4.1 can be refined according to further
conditions on $\Phi$ and $\gamma$. In the examples to follow, based on
homogeneous spaces, there is a natural such condition (see
Proposition~\ref{homog.bundle.gen}
below). On the other hand we feel that the right direction for a general
formulation should be preceded by still further examples than these. We will
not attempt this here.

\section{Examples}

In this section we come to the main task of the paper, which is to construct
concrete
examples of non-trivial quantum principal bundles and connections on them. This
justifies the
formalism developed in the last section and in Appendix A. We begin with a
general development of
quantum homogeneous spaces, both with universal and non-universal calculi. This
includes the trivial frame bundle of $S^3$ in a non-commutative setting based
on the quantum double Hopf algebra. We then give the full details of the
simplest case of our construction where the homogeneous space is a $q$-deformed
$S^2=SO(3)/U(1)$ and the canonical connection on the associated bundle is a
q-deformed Dirac monopole. This application is perhaps the main result of the
paper and demonstrates in detail the various assumptions and theorems above and
their smooth classical limit to the usual geometry as $q\to 1$.

\subsection{Bundles on quantum homogeneous spaces}

We begin with the simplest example of all, namely with trivial base and
connection given by the Maurer-Cartan form. This provides a useful warm-up for
quantum homogeneous spaces as well as an instructive look at the content of our
various axioms. We consider for our quantum principal bundle the base $B=k$,
total space is $P = A$ and the trivialization $\Phi$ given by the identity map.
Recall here that every Hopf algebra coacts on itself by the right regular
coaction provided by the coproduct $\Delta$. We suppose also that the
differential structure on $P$ is taken to be
the same as that on $A$.

\begin{ex} Let $P=A$ be a Hopf algebra equipped with the bicovariant
differential calculus defined by an ideal $M_P=M_A$ in $\ker\eps$. Let
$\Delta_R=\Delta$ be the right regular coaction. Then $P(k,A,M_A)$ is a trivial
quantum principal bundle in the sense of Example~\ref{trivial.bundle.gen} with
trivialization $\Phi=\id$. The bundle is equipped with a trivial connection
$\Pi=\id$ with $\Gamma_{ver}=\Gamma_P$ and corresponding connection 1-form
\begin{equation}
\omega:A\to \Gamma_P,\qquad \omega (a) = \sum (Sa_{(1)}) da_{(2)}.
\end{equation}
This is the Maurer-Cartan form on the Hopf algebra $A$.
\end{ex}
\proof That this obeys conditions 1.-4. in
Definition~\ref{principal.bundle.gen} is elementary. For condition 3. we have
only to note that because the coproduct has a counit $\eps$, it follows that if
$\Delta(b)=b\tens 1$ then  $b=\eps(b) 1$. Hence condition 3. holds with $B=k$.
The freeness condition 4. follows because $A$ has an antipode $S$ so that
$(\cdot\tens\id)(\id\tens\Delta)(\sum a Sb\o\tens b\t)=\sum a\tens b$. This is
the content of the linear isomorphism $\kappa$ in (\ref{kappa.A}).
Equivalently, the existence of the antipode $S$ is precisely the requirement
that $\Phi=\id$ is convolution-invertible as needed in
Example~\ref{trivial.bundle}. Its is clearly also an intertwiner and hence a
trivialization, from which both freeness and exactness  follow from
Example~\ref{trivial.bundle} in the universal case. In the non-universal case
we note that the covariance condition 5. $\Delta(N_A)\subset N_A\tens A$ is
just the condition that the differential calculus defined with ideal $M_A$ and
corresponding sub-bimodule $N_A$ is left-covariant, as explained in
Section~4.2. Finally, the equality $\widetilde{\ }(N_A)=A\tens M_A$ follows
using
again the linear isomorphism $\kappa:A\tens A\to A\tens A$. In this case the
exactness condition follows from Example~\ref{trivial.bundle.gen}.

Since $\Gamma_{hor} = 0$ there is a natural (trivial)
connection $\Pi$ in $P$, given by $\Pi (\rho) = \rho$
for any $\rho \in \Gamma_P = \Gamma_A$. From
Proposition~\ref{prop.connection.gen} we know that it has
a connection 1-form, which one can compute as shown. To also see directly that
$\omega $ is covariant
under the adjoint coaction $Ad_{R}$, we have
\begin{eqnarray*}
\Delta_{R}\omega(a) & = & \sum \Delta (Sa_{(1)}) (d \otimes id) \Delta
(a_{(2)}) = \sum
Sa_{(2)} da_{(3)} \otimes (Sa_{(1)}) a_{(4)} = (\omega \otimes id)
\Ad_R (a)
\end{eqnarray*}
for any $a \in A$. Condition~3 in Proposition~\ref{prop.connection.gen} holds
from the map $\kappa$ in (\ref{kappa.A}). It is related to the
$Ad_R$-invariance of $M_A$ arising from the assumption that the differential
calculus is bicovariant. Condition 1. is also easy. From another point of view,
the trivialization $\Phi$ in this case obeys the condition (\ref{phi.nice})
sufficient to define a basis-independent trivial connection
$\Phi^{-1}(a\o)d\Phi(a\t)$, which is $\omega$. These considerations are of
course unecessary  for the universal differential calculus where $M_A=\{0\}$.
\endproof

Thus the various points of view in the theory of Section~4 manifestly tie up in
this example.
Next let us assume that $P$ is a quantum group such that there is an Hopf
algebra projection  $\pi : P \rightarrow A$. (This corresponds in the
classical case to an inclusion of groups $G\subseteq P$ say). The
right regular coaction of $P$ on itself pushes out by $\pi$ to a
coaction $\Delta_R=(\id\tens\pi)\circ\Delta:P\to P\tens A$ and we
define the associated {\em quantum homogeneous space} as:
\[ B = P^A \equiv \{ b \in P : \sum b_{(1)} \otimes \pi (b_{(2)}) = b
\otimes 1 \}. \]
In the classical situation there is a principal bundle over the underlying
classical homogeneous space. A
theorem of Chevalley ensures that the bundle is locally trivial in the usual
sense. Later we will give a criterion for patching in the quantum case, but for
now we concentrate on the global properties expressed in
Definitions~\ref{principal.bundle} and~\ref{principal.bundle.gen}.
A useful sufficient condition for a bundle is

\begin{lemma}\label{homog.bundle} Let $\pi:P\to A$ be a Hopf algebra map and a
surjection between two Hopf algebras $A,P$. Let $\Delta_R$ be the induced
coaction by pushout of $\Delta$ and $B=P^A$. If $\pi$ is such that
\[ \ker\pi\subset \cdot(\ker\pi|_B\tens P)\]
then $P(B,A,\pi)$ is a quantum principal bundle in the sense of
Definition~\ref{principal.bundle} with the universal differential calculus. We
say that $\pi$ obeying this assumption is {\em exact}.
\end{lemma}
\proof Since $\pi$ is a surjection, freeness of the induced coaction $\Delta_R$
follows at once from freeness of the right coaction in the preceding example.
We use that $P$ is a Hopf algebra. In the universal case it remains to prove
the exactness condition~5 in Definition~\ref{principal.bundle}. This needs some
condition on $\pi$ and a convenient one for our applications is as stated. Note
that
$\pi=\eps$ when restricted to the fixed subalgebra $j(B)\subset P$.
Assuming the condition let $\rho\in P^2$. From the linear isomorphism $\kappa:
P\tens P\to P\tens P$ in (\ref{kappa.A}) applied to the Hopf algebra $P$ we can
write $\rho=\sum\kappa(w^k\tens u^k)$ for $u^k\in \ker\eps$ and $w^k\in P$ with
the latter set linearly independent. Then $\widetilde{\
}\rho=(\id\tens\pi)\circ\kappa^{-1}\rho=\sum w^k\tens\pi(u^k)$ and hence if
$\rho\in\ker\widetilde{\ }$ we conclude that $\pi(u^k)=0$. For each of these,
we can write from our assumption on $\pi$ that $u^k=\sum_i b^k{}_iv^k{}_i$
where $b^k{}_i\in\ker\eps|_B$ and $v^k{}_i\in P$. Then
\align{\rho&=&\sum w^k(Su^k\o d u^k\t)=\sum w^k
(Sv^k{}_i\o)(Sb^k{}_i\o)d(b^k{}_i\t v^k{}_i\t)\\
&=&\sum \eps(b^k{}_i)w^k
(Sv^k{}_i\o)dv^k{}_i\t+w^k(Sv^k{}_i\o)(Sb^k{}_i\o)(db^k{}_i\t)v^k{}_i\t}
using the Leibniz rule in $\Gamma_P$. The first term vanishes by our assumption
and the second term lies in $\Gamma_{hor}$. Hence $\ker\widetilde{\
}=\Gamma_{hor}$ as required. \endproof

Next we come to the construction of connections. We recall for classical
homogeneous spaces that
in the compact semisimple case there is a canonical
connection on the bundle. It is defined by an $ad$-invariant splitting
of the Lie algebra $\goth p=\goth m\oplus \goth g$ (provided by the
Killing form). See \cite{KobNom:fou}. Such a splitting can be viewed
as inducing a coalgebra (but not usually algebra) map $U(\goth p)\to
U(\goth g)$ covering the inclusion $U(\goth p)\supseteq U(\goth g)$
(the map sets $\goth m$ to zero). In our dual quantum group
formulation then this means an algebra but not usually coalgebra map
$i:A\to P$ which is $Ad$-covariant in a suitable sense and which obeys
$\pi\circ i=\id.$
We assume this data now for our quantum homogeneous space.

\begin{prop}\label{homog.connection}
Let $P(B,A,\pi)$ be quantum principal bundle over a homogeneous space
and with
universal differential structure.
If there is an algebra map $i:A\to P$ such that $\pi\circ
i=\id$, $\epsilon (i(a)) = \epsilon (a)$ for any $a \in A$, and
\[ (id \otimes \pi )Ad_R i = (i \otimes id )Ad_R .\]
then
\[ \omega (a) = \sum Si(a)_{(1)} di(a)_{(2)} \]
is a connection 1-form. We call the corresponding $\Pi$ from
Proposition~\ref{prop.connection} the {\em canonical connection} on the quantum
homogeneous space.
\end{prop}
\proof We have to check that $\omega$ obeys the assumptions of
Proposition~\ref{prop.connection}. First we prove
that $\omega$ is  $Ad_{R}$-covariant,
\begin{eqnarray*}
\Delta_R \omega (a) & = & \sum Si(a)_{(2)} di(a)_{(3)} \otimes \pi
(Si(a)_{(1)}) \pi \circ i (a)_{(4)} \\
& = & \sum Si(a_{(2)})_{(1)} di(a_{(2)})_{(2)} \otimes (Sa_{(1)})a_{(3)} \\
& = & \sum \omega (a_{(2)}) \otimes (Sa_{(1)})a_{(3)}
\end{eqnarray*}
where in the second equality we used the fact that $i$ is an
intertwiner of $(\id\tens\pi)Ad_R$ on $P$ and $Ad_R$ on $A$ as in the
hypothesis.

Next we apply the map $\widetilde{\ }$ to $\omega$ to obtain
\begin{eqnarray*}
\widetilde{\ } \omega(a) & = & \sum (Si(a)_{(1)}) i(a)_{(2)} \otimes \pi
(i(a)_{(3)}) - \sum (Si(a)_{(1)})i(a)_{(2)} \otimes 1 \\
& = & 1 \otimes \pi (i(a)) - \epsilon (a) \otimes 1 = 1 \otimes a -
\epsilon (a) \otimes 1.
\end{eqnarray*}
We now apply Proposition~\ref{prop.connection} to conclude the result.
\endproof

\begin{cor} Let $P{{\buildrel \pi\over\to}\atop{\hookleftarrow\atop
i}}A$ be a Hopf algebra projection, i.e. suppose that $i$ is a Hopf
algebra map and covered by $\pi$. This is an example of a quantum homogeneous
space with universal differential calculus as in
the preceding proposition. The bundle is trivial with trivialization
given by $i$ itself. The canonical connection $\omega$ above then
coincides with the flat connection in Example~\ref{trivial.connection}.
\end{cor}
\proof Because $i$ is assumed to be a Hopf algebra map, and
$\pi\circ i=\id$, it is immediate that it is an intertwiner for
$\Delta_R$ on $A$ and $P$, and therefore defines a trivial bundle $P(B,A,i)$
from Example~\ref{trivial.bundle}.
One can also go through Lemma~\ref{homog.bundle} which is satisfied in this
trivial
case. The map $i$ is also covariant for $\Delta_L$ and hence $Ad$-invariant in
the
way required in Proposition~\ref{homog.connection}. Hence we can apply that
proposition to obtain a connection. We note that Hopf algebra projections of
the type that we have assumed here
are familiar in the theory of Hopf algebras\cite{Rad:str}\cite{Ma:skl}, where
it is known that $P$ here is necessarily isomorphic to a semidirect
product, $B\cocross A\isom P$. This is built on the
linear space $B\tens A$ with cross relations according to the action
\[ a \triangleright b = \sum i(a_{(1)})b i (Sa_{(2)})\]
and gives the explicit structure of the trivial bundle in this case.
 \endproof

\note{The map $\theta:B\cocross A:b\tens a\mapsto j(b)i(a)$ has inverse
 \[ \theta^{-1} (u) \equiv \sum u_{(1)} (i \circ \pi)(S(u_{(2)}))
\otimes \pi (u_{(3)}). \]
For this formulation see \cite[Prop. A.2]{Ma:skl}, where it is also noted that
$B$ is a Hopf algebra living in the braided category of
$D(A)$-modules. Here $D(A)$ is Drinfeld's quantum double Hopf algebra
of $A$.
For completeness we verify for ourselves the most important of the
facts needed here.
Firstly, $\theta^{-1}(u) \in B \otimes A$ for any $u \in P$ because
\begin{eqnarray*}
(\Delta_{R} \otimes id)\theta^{-1}(u) & = & \sum u_{(1)(1)} (i \circ
\pi (Su_{(2)}))_{(1)} \otimes \pi (u_{(1)(2)} (i \circ \pi
(Su_{(2)}))_{(2)}) \otimes \pi (u_{(3)}) \\
& = &\sum u_{(1)(1)} (i\circ \pi (Su_{(2)}))_{(1)} \otimes \pi
(u_{(1)(2)} \pi (Su_{(2)})_{(2)}) \otimes \pi (u_{(3)}) \\
& = & \sum u_{(1)} (i \circ \pi )(Su_{(2)}) \otimes 1 \otimes \pi
(u_{(3)})
\end{eqnarray*}
In second equality we used that $\pi$ is an algebra map and that $i$ is a Hopf
algebra
map, while in the third one we used that $\pi$ is a Hopf algebra map. To
complete the proof we need to
check if $\theta^{-1}$ is inverse of $\theta$. We have
\begin{eqnarray*}
\theta \circ \theta ^{-1} (u) & = & \sum \theta (u_{(1)} (i \circ
\pi)(Su_{(2)}) \otimes \pi (u_{(3)})) = \sum j(u_{(1)} (i \circ \pi
)(Su_{(2)}))(i\circ \pi)(u_{(3)}) \\
& = & \sum u_{(1)} (i \circ \pi (Su_{(2)}u_{(3)})) = u.
\end{eqnarray*}
and similarly on the other side.}

This corollary provides an important source of (trivial) quantum bundles.

\begin{ex} Let $A$ be a finite-dimensional quasitriangular Hopf
algebra in the sense of \cite{Dri}. This means that it is equipped
with an element $\CR\in A\tens A$ obeying some axioms. Let $P=D(A)$ be
the quantum double of $A$ as a Hopf algebra  built on the linear space
$A^*\tens A$ \cite{Dri}. It is known that there is a Hopf algebra
projection\cite{Ma:dou}
\[ D(A){{\buildrel \pi\over\to}\atop{\hookleftarrow\atop i}}A,\quad
\pi(\phi\tens a)=(S\phi \otimes id) (\CR) a,\quad i(a)=1\tens a.\]
where $\CR=\sum\CR\uo\tens\CR\ut$. Hence $P=D(A)$ is a trivial quantum
principal bundle with structure quantum group $A$. It was also shown in
\cite{Ma:dou} that we can identify the base $B=P^A$ as the algebra
\[ B=A^*, \quad b\und\cdot c=\sum b\t c\th <\CR,b\th\tens Sc\o)<\CR, b\o\tens
c\t>,\qquad \forall b,c\in A^*\]
where the right hand side expresses the product of $B$ in terms of that of
$A^*$. The corresponding element of $P$ is $j(b)=\sum b\o<\CR\uo,b\t>\tens
\CR\ut$.
\end{ex}
\proof  We use here the conventions in which the $D(A)$ has the tensor
product comultiplication and a certain double-semidirect product algebra
structure. The structure of $B$ here is that of the braided group of
function algebra type associated to the dual quantum group
$A^*$\cite{Ma:eul}. Note that $A$ here is of enveloping algebra type
(a quasitriangular Hopf algebra) being regarded perversely as a
`functions' on some dual group. With this description of $B$ the map
$\theta:B\cocross A\isom D(A)$ is\cite[Prop. 4.1]{Ma:skl} (where $A$ is denoted
$H$),
\[ \theta(b\tens a)=\sum b\o <\CR\uo,b\t>\tens \CR\ut a=j(b)i(a)\]
for $j$ as stated and the fact $(b\tens a)(1\tens a')=b\tens aa'$ for
the product in $D(A)$. \endproof

The base of this bundle then is the algebra $B$ introduced in
\cite{Ma:eul} in another context. It is (in a certain sense) a
braided-commutative Hopf
algebra living in the braided category of
$A$-modules. We do not discuss it further except to note that the
example of $B$ when $A=U_q(sl_2)$ is computed in \cite{Ma:eul} and
called $BSL_q(2)$.
 Just as $SU_q(2)$ is some kind of quantum 3-sphere, $BSL_q(2)$
equipped with a suitable
$*$-algebra structure (which exists) can be called a {\em braided
3-sphere}\cite{Ma:eul}. This is the base for this case of the construction.
Since $A=U_q(sl_2)$ is being regarded as one of function
algebra type, the `underlying' structure group in this case should be
thought of as some kind of deformation of a dual of $sl_2$. Of course, the
algebras and Hopf algebras here are not finite-dimensional so appropriate care
has to be taken to work with the correct generators.

The simplest case of the preceding construction is when $A=kG$ is
the group algebra of a finite group $G$. This is quasitriangular with
$\CR=1\tens 1$. In this case $D(G)=k(G){}^{Ad}\cocross kG$. Here $B=k(G)$ so
that the base is classical, namely the discrete group $G$. The fiber
on the other hand has structure group $kG\isom k(\hat G)$ in the case
where $G$ is Abelian. Here $\hat G$ is the character group of $G$ and
forms the classical structure group of our bundle. When $G$ is non-Abelian
there is no
such group $\hat G$. Instead, we can continue to do gauge theory with
the non-commutative algebra $kG$ in place of functions on $\hat G$.
This is a typical application of non-commutative geometry to groups.

We can also dualize the above construction to obtain a different
bundle. This time we begin with a finite-dimensional quasitriangular
Hopf algebra $(H,\CR)$ with $\CR\in H\tens H$. $D(H)^*$ is the dual
Hopf algebra of Drinfeld's double. It has as algebra structure the
tensor product algebra $H\tens H^*$, but a doubly-twisted coalgebra
structure. This works out \cite[Appendix]{Ma:skl} as
\[ \Delta h\tens a=\sum h\t\tens (S f^b\o)a\o f^b\th\tens e_b\tens
a\t<f^b\t,h\o>,\qquad \eps(h\tens a)=\eps(h)\eps(a)\]
where $\{e_b\}$ is a basis of $H$ and $\{f^b\}$ a dual basis.

\begin{ex} Let $H$ be a finite-dimensional quasitriangular Hopf algebra
and $A=H^*$ its
dual. Let $P=D(H)^*$ as described. Then
\[ P{{\buildrel \pi\over\to}\atop{\hookleftarrow\atop i}} A,\qquad
\pi(h\tens a)=\eps(h)a,\quad i(a)=<a\o,\CR\ut>S\CR\uo\tens a\t\]
is a Hopf algebra projection as in the above corollary and hence defines a
quantum principal bundle on a quantum homogeneous space. The base $B$
can be identified as $B=H$ (as an algebra). The map $j$ is then $j(b)=b\tens
1_A$.
\end{ex}
\proof This is obtained by dualizing the preceding example in an
elementary way. The maps $\pi,i$ in the preceding example dualise to
the maps $i,\pi$ respectively now. The base $B$ also has a
braided-coalgebra structure (making it a braided group) though this
need not concern us now. \endproof

Some examples of this dual quantum double have been studied in
\cite{PodWor:def} as $C^*$-algebras, so many of the details here for
an operator-algebraic treatment are already known. The double in the
case when $H=U_q(sl_2)$ or more precisely, $A=SL_q(2)$ (with a
suitable $*$-structure) is called the
quantum Lorentz group. Moreover, because $H$ here is a factorizable
quantum group one can show that $U_q(sl_2)\isom BSL_q(2)$ as algebras
\cite[Cor. 2.3]{Ma:skl} (for generic $q\ne 1$).
Thus we see that the quantum Lorentz group is a trivial bundle with
$SL_q(2)$ fiber and a base which is again our braided-$S^3$. It seems
reasonable to view this trivial bundle
\[ P(BSL_q(2),SL_q(2)){{\buildrel
\pi\over\to}\atop{\hookleftarrow\atop i}} SL_q(2)\]
with appropriate $*$-structures as a kind of {\em frame bundle} for our
braided-$S^3$. The flat
connection $\omega$ in this case should be thought of as the quantum
spin-connection corresponding to its parallelization.

Finally, in the case when $H=kG$, the fiber is the classical (albeit,
discrete) group $G$ and the base is $\hat G$ in the Abelian case,
viewed as a non-commutative space in the non-Abelian case.

This completes our construction at the level of universal differential calculus
and some examples.
The ones constructed via the
corollary have trivial bundles and hence flat (and other) connections
on them. Next, we come to the corresponding refinements for the non-universal
case.

\begin{prop}\label{homog.bundle.gen} For $(P,A,\pi)$ as in
Lemma~\ref{homog.bundle} we suppose further that $P$ is equipped with a
left-covariant differential structure generated by a right-ideal $M_P$, and $A$
with a bicovariant one with ideal $M_A$. If
\begin{enumerate}
\item $(id\otimes\pi)Ad_R(M_P)\subset M_P\otimes A$
\item $M_A = \pi (M_P)$
\end{enumerate}
then $P(B,A,\pi,M_P,M_A)$ is a quantum principal bundle in the sense of
Definition~\ref{principal.bundle.gen}.
\end{prop}
\proof We have to prove the conditions 5-7 in
Definition~\ref{principal.bundle.gen}. The last of these builds on the
exactness already proven in the universal case. First we prove covariance under
$A$. Thus our first condition  implies that for any $v\in M_P$ we have
\[\sum\kappa (1\otimes v\t)\otimes\pi((Sv\o)v_{(3)}) = \sum Sv\t\tens
v_{(3)}\tens \pi((Sv\o)v_{(4)})\in N_P\tens A \]
where $(\id\otimes\pi)Ad_R(v)=\sum v\t\tens\pi((Sv\o)v_{(3)})$ in an explicit
notation. Consequently
for any $u\in P$ we have
\[\sum u\o Sv\t\tens v_{(3)}\tens \pi(u\t(Sv\o)v_{(4)})\in N_P\tens A. \]
Let $\rho = \sum\kappa (u^k\tens v^k)\in N_P$, where $u^k\in P$ and
$v^k\in M_P$. Then
\[\Delta_R\rho = \Delta_R(\sum u^kSv^k\o\tens v^k\t) = \sum
u^k\o Sv^k\t\tens v^k_{(3)}\tens\pi(u^k\t(Sv^k\o)v^k_{(4)})\in N_P\tens A
\]
as required for the covariance condition~5 in
Definition~\ref{principal.bundle.gen}.
Meanwhile, our second condition for $M_A$ combined with the observation
$\kappa^{-1}(N_P) = P\otimes M_P$ and
$\widetilde{\  } = (id\otimes \pi)\kappa^{-1}$ gives the condition~6 in
Definition~\ref{principal.bundle.gen} for the projection of $\widetilde{\ }$
down to a map $\widetilde{\ }_{N_P}$. Finally, we need the exactness
condition~7 with respect to this map. We write any representative $\rho_U\in
P^2$ of $\rho\in\ker\widetilde{\ }_{N_P}$ in the same way as in the proof of
Proposition\ref{homog.bundle} and this time have $\sum w^k\tens\pi_A\pi(u^k)=0$
and hence $\pi(u^k)\in M_A$. Here $\pi_A$ is the canonical projection to
$\ker\eps/M_A$ for the kernel of the counit of $A$. Then from our
second condition
on $M_P$ we know there exist $u'{}^k\in M_P$ with $\pi(u^k-u'{}^k)=0$.
Moreover, $\rho'_U=\sum\kappa(w^k\tens (u^k-u'{}^k))$ has the same image $\rho$
in $\Gamma_P$ but now lies in $\ker\widetilde{\ }$. Hence by
Lemma~\ref{homog.bundle} we conclude that $\rho\in \Gamma_{hor}$.
\endproof

\begin{prop}\label{homog.connection.gen}
Let $P(B,A,\pi,M_P,M_A)$ be a quantum principal bundle over the homogeneous
space
$B$ equipped with a differential structure as in
Proposition~\ref{homog.bundle.gen}. If there is an algebra map
$i:A\to P$ obeying the hypothesis of Proposition~\ref{homog.connection} and in
addition
\[ i(M_A)\subset M_P\]
then
\begin{equation}
\omega(a) = \sum Si(a)\o di(a)\t.
\end{equation}
defines a connection 1-form. We call the corresponding connection $\Pi$ from
Proposition~\ref{prop.connection.gen} the canonical connection.\end{prop}
\proof Now we show that the map $\omega$ satisfies the hypothesis of
Proposition~\ref{prop.connection.gen}. First, $\omega(1) = 0$ because
$i$ is an algebra map. Let us denote by
$\pi_{N_P}:P^2\rightarrow\Gamma_P$ a canonical epimorphism. Then we have
\begin{eqnarray}
\omega(a) & = & \sum Si(a)\o di(a)\t = \sum \pi_N(Si(a)\o(1\otimes
i(a)\t - i(a)\t\otimes 1) \nonumber \\
& = &\sum \pi_{N_P}(Si(a)\o\otimes i(a)\t) = \pi_N\kappa(1\otimes i(a)).
\label{omega.kappa}
\end{eqnarray}
If $a\in M_A$ then $i(a)\in M_P$, and $\kappa(1\otimes i(a))\in N_P$.
Therefore $\omega (a) = 0$ for any $a\in M_A$. Similarly to the
proof of Proposition~\ref{homog.connection}
 we can show that
\[\widetilde{\  }_{N_P} \circ \omega(a) = 1\otimes \pi_A(a-\eps(a)).\]
Finally the map $\omega$ is
$Ad_R$-covariant by the same argument as in
the proof of Proposition~\ref{homog.connection}.
Applying Proposition~\ref{prop.connection.gen} we obtain the assertion.
\endproof

It is obvious from this that if $i$ is a Hopf algebra map then the bundle is
trivial with trivialization $\Phi=i$ and the canonical connection is then the
trivial one associated to this (here $\Phi$ is a Hopf algebra map and obeys the
condition (\ref{phi.nice}) so that there is a unique trivial connection).
Rather more useful for us in the next section is a kind of `local' form of
Proposition~\ref{homog.connection.gen} as follows. We suppose for this that
$P(B,A,\pi)$
is a locally trivial quantum principal bundle over the
homogeneous space $B$ in the sense that we are given one or more trivial
bundles
$P_k(B_k,A,\pi_k)$ of the type above and inclusions $P\to P_k$ etc as at the
end of Section~4.1, which we suppose now to be compatible with the $\pi_k$ in
the obvious sense.

\begin{prop}\label{homog.connection.local}
Let $P(B,A,\pi)$ be locally trivial  with trivial bundles $P_k(B_k,A)$ as
explained.
Let $\{\omega^i \}$ denote a basis of left-invariant
differential forms for $\Gamma_P$ and assume that $\Gamma_{P_k} = P_k\{
\omega^i\}$ is the differential structure on each $P_k(B_k,A)$. In this
situation if for one of these
$P_k(B_k,A)$ there exists an $Ad_R$ covariant map $i: A\hookrightarrow
P_k$ such that $\pi\circ i = id$ on $P_k$ then
the map $\omega(a) = \sum Si(a)\o di(a)\t$ is globally defined on $P$ and
defines a connection $\Pi$.
\end{prop}
\proof We have to show the map $\omega$ is defined globally. The rest
of the proposition is deduced from Proposition~\ref{homog.connection.gen}.
We represent $\omega(a)$ in the basis of the
left-invariant one-forms $\{\omega^i\}$. Let $\chi_i\in P^*$ be such
that\cite{Wor:dif}
\[du = \sum u\o\chi_i(u\t)\omega^i\]
for any $u\in P$. Using this representation we find
\begin{eqnarray}
\omega(a) & = & \sum (Si(a)\o)i(a)\t \chi_i(i(a)_{(3)})\omega^i
\nonumber \\
& = & \sum \chi_i(i(a))\omega^i. \label{chi.omega}
\end{eqnarray}
Because $\chi_i(i(a))$ are defined for each $a$ and $\omega^i \in
\Gamma_P$, the map $\omega$ is defined globally. \endproof

\subsection{Dirac monopole bundle and its canonical connection}

We now come to  the explicit construction of
a non-trivial bundle by the general methods introduced above. This is a
$q$-deformed analog of the usual Dirac
$U(1)$ connection on $S^2$ obtained as the canonical
connection in Proposition~\ref{homog.connection.gen} with $P=SO_q(3)$ and a
suitable differential calculus. The base in this case a $q$-sphere in the sense
of \cite{Pod:sph} and our construction has a smooth limit as $q\to 1$ to the
usual Dirac monopole and its connection (with the usual classical differential
calculus). This serves as an important check on our constructions, as
well as providing a novel Hopf-algebraic derivation of this important
configuration. We first construct the bundle for any suitable calculus
(including the universal calculus as in Section~4.1) and then specialise to the
3-D calculus of Woronowicz\cite{Wor:dif} for the computation of the connection.

For the standard construction of a monopole one works with $S^2$ as the
homogeneous space $Spin(3)/Spin(2)=SU(2)/U(1)$. The canonical connection on
this is the monopole of charge one. One can also take
 $S^2=SO(3)/U(1)$ where the previous $U(1)$ is a double cover of the new $U(1)$
and we arrive at a monopole of charge two. We will construct the quantum
version of the second case, but will discuss both as far as possible.
We begin by developing the classical theory in the algebraic setting above. Of
course, we work with
the functions on
$SU(2)$ and $SO(3)$ rather  than points themselves. Generating the functions
on the former are
the matrix  co-ordinate functions $\pmatrix{\alpha&\beta\cr
\gamma&\delta}$ where  $\alpha(X)=X^1{}_1$ etc for a matrix $X\in
SU(2)$. They obey  the relations of commutativity and
$\alpha\delta-\beta\gamma=1$.

Next there is a canonical inclusion of $U(1)$ in $SU(2)$ along the diagonal. In
algebraic terms this is given by a
projection
\[ \pi\pmatrix{\alpha&\beta\cr \gamma&\delta}=\pmatrix{Z^\h
&0\cr 0&Z^{-\h}}\]
where $A=k[Z^\h,Z^{-h}]$ is algebra of functions on $U(1)$.  The matrix
comultiplication on $SU(2)$ is $\Delta
\alpha=\alpha\tens\alpha+\beta\tens\gamma$ etc, and this  induces a
coaction of $k[Z^\h,Z^{-\h}]$ via
\[\Delta_R\pmatrix{\alpha&\beta\cr
\gamma&\delta}=\pmatrix{\alpha\tens Z^\h&\beta\tens  Z^{-\h}\cr
\gamma\tens Z^\h &\delta\tens Z^{-\h}}.\]
This extends to products as an algebra homomorphism (a comodule algebra) as
required for the general theory. For example
$\alpha\beta\mapsto\alpha\beta\tens 1$, $\alpha\gamma\mapsto\alpha\gamma\tens
Z$ etc.
{}From this it follows that the algebra of functions on the sphere is then the
fixed-point subalgebra $B$ of
$SU(2)$ with generators
\[ B=SU(2)^{k[Z^\h,Z^{-\h}]}=<1,b_-=\alpha\beta, b_+=\gamma\delta,
b_3=\alpha\delta>\]
and $b_-b_+=b_3(1-b_3)$.

Note that these algebras are $*$-algebras. The relations
$\alpha^*=\delta,\beta^*=-\gamma$ imply that $b^*_\pm=-b_\mp$ while
$b_3^*=b_3$. Writing $b_\pm=\pm(x\pm\imath y)$ and $z=b_3-\h$ it is easy
to see that the algebra $B$ describes a sphere of radius $\h$ in the
usual Cartesian
co-ordinates. Next, assuming that $b_3\ne 0$, every remaining element
of $SU(2)$
can be written uniquely in the form
\[ \pmatrix{\alpha&\beta\cr \gamma&\delta}=\pmatrix{\sqrt{b_3}&
{b_-\over\sqrt{b_3}}\cr
{b_+\over\sqrt{b_3}}&\sqrt{b_3}}\pmatrix{e^{\imath\theta}&0\cr
0&e^{-\imath\theta}}\]
which gives one co-ordinate chart of $SU(2)$. The corresponding
fiber co-ordinate function that returns the $U(1)$ group co-ordinate
$e^{\imath\theta}$ is
\[ \Phi_0(Z^\h)=\sqrt{\delta^{-1}\alpha}
,\qquad\Phi_0(Z^{-\h})=\sqrt{\alpha^{-1} \delta}.\]
There is another co-ordinate chart that works when $1-b_3\ne 0$,
\[ \pmatrix{\alpha&\beta\cr \gamma&\delta}=\pmatrix{
{b_-\over\sqrt{1-b_3}}&\sqrt{1-b_3}\cr
-\sqrt{1-b_3}&{b_+\over\sqrt{1-b_3}}}\pmatrix{ e^{\imath\phi}&0\cr
0&e^{-\imath\phi}}.\]
The corresponding fiber co-ordinate function is
\[ \Phi_1(Z^\h)=\sqrt{-\gamma\beta^{-1}},\quad
\Phi_1(Z^{-\h})=\sqrt{-\beta\gamma^{-1}}.\]
These can be used to give trivial bundles over the relevant patches. Over $\C$
there is no problem with the square roots here. On the other hand they will
be problematic in the general algebraic case and for this reason we pass now to
the charge two setting with $SO(3)$.

To work with $SO(3)$ we note that because the relations of $SU(2)$
are either
homogeneous or change  degree by $2$, there is an automorphism of the
algebra of functions given  by $\pmatrix{\alpha&\beta\cr
\gamma&\delta}\mapsto \pmatrix{-\alpha&-\beta\cr -\gamma&-\delta}$.  The
fixed point subalgebra under  this automorphism is (the algebra of
functions on)
$SO(3)$ and consists
 precisely of expressions of even degree, i.e. is generated by
$<1,\alpha\beta,\alpha\gamma,\cdots>$  as a subalgebra of the
functions on $SU(2)$. The same applies in the quantum case below.
For the structure group one has to work with a different but isomorphic $U(1)$
to the one above. In our function algebra
language one has to work with $A=k[Z,Z^{-1}]$ as a sub-Hopf algebra of the one
above. Clearly the fixed subalgebra $B$ in $SO(3)$ by this sub-Hopf algebra is
just the same as the fixed subalgebra above. This is because the generators of
the latter are already of even degree.

With this description of the function algebra of $SO(3)$ the corresponding
co-ordinate chart for $b_3\ne 1$ comes out now as
\[ \Phi_0(Z)=\delta^{-1}\alpha,\qquad\Phi_0(Z)=\alpha^{-1} \delta\]
and for $1-b_3\ne 0$ as
\[ \Phi_1(Z)=-\gamma\beta^{-1},\quad \Phi_1(Z)=-\beta\gamma^{-1}.\]
The first gives a trivialization of the
bundle $P_0=SO(3)[\delta^{-1}\alpha,\alpha^{-1}\delta]$ over
$B_0=B[b_3^{-1}]$, and the second of the bundle
$P_1=SO(3)[\gamma\beta^{-1},\beta\gamma^{-1}]$ over
$B_1=B[(1-b_3)^{-1}]$, in both cases with structure Hopf algebra
$k[Z,Z^{-1}]$. Note that we are restricting to functions in open sets
$b_3\ne 0$ etc by means of localization. Finally, there is a bundle $P_{01}$
over
$B_{01}$ obtained by making both localizations simultaneously. One may
check that these are all trivial bundles (so $P_0=B_0k[Z,Z^{-1}]$ etc.)
and that the maps are intertwiners for $\Delta_R$ and the right
regular coaction of $k[Z,Z^{-1}]$ on itself. Finally, they
paste-together correctly because the ratio
\[\gamma_{01}(Z)=\Phi_0(Z)\Phi_1(Z)^{-1} =
-\delta^{-1}\alpha\beta\gamma^{-1}
=- b_+^{-1}b_- = -((b_3-1)b_3)^{-1} b_-^{2}\]
lies in $B_{01}$ as it should.

For the canonical connection on this bundle, we look for an $Ad$-covariant
algebra map $i:k[Z,Z^{-1}]\to
SO(3)$ to use in Proposition~\ref{homog.connection.gen}.
 Since the $Ad_R$ action of $k[Z,Z^{-1}]$ on itself is
trivial, $i(Z)$ must be a $(\id\tens\pi)Ad_R$-invariant element of the function
algebra $SO(3)$.
Computing this gives that it must be a
combination of $\alpha,\delta$. We arrange $\pi\circ i=\id$ if we take
$i(Z)=\delta^{-1}\alpha$. Note that this does not exist globally,
indeed it coincides with the co-ordinate chart $\Phi_0$. But from
Proposition~\ref{homog.connection.local} we know
that the resulting connection $\omega$ is globally defined provided
differential structures on $P_0$ and $P$ are generated by the same
ideal $M_P\subset\ker\eps$.
For now we proceed locally, concentrating on this co-ordinate chart. A
further complication caused by this is that $P_0$ is only a formal Hopf
algebra (the comultiplication $\Delta (\delta^{-1}\alpha )$ is a formal
power-series). Again, this does not affect the answer.

\begin{prop}\label{class.monopole} Applying
Proposition~\ref{homog.connection.gen} to the bundle $P_0$
over $B_0$, the  map $i$, and the classical differential calculus $d$,
we find that  the canonical connection
$$\omega(Z)=\sum Si(Z)\o d i(Z)\t$$
exists globally and equals the Dirac $U(1)$ monopole
connection of charge two,
\[ \omega(Z)=\cases{\beta_0(Z)+\Phi_0^{-1}(Z)d\Phi_0(Z),\quad
\beta_0(Z)={b_+ d b_--b_-db_+\over b_3}= 2\imath{(xdy -ydx) \over z+\h}&\cr
\beta_1(Z)+\Phi_1^{-1}(Z)d\Phi_1(Z),\quad \beta_1(Z)={b_+ d b_--b_-db_+\over
b_3-1}=2\imath{(xdy-ydx)\over z-\h}&}.\]
\end{prop}
\proof The formal proof that the ideals $M_P$ etc defining the usual
commutative calculus obey the
relevant conditions will follow immediately from
Proposition~\ref{prop.connection.3d} (by setting $q=1$) so we do not give this
separately here. It is however, quite instructive to compute
 $\omega$ from Proposition~\ref{homog.connection.gen} and see that
it gives (a new algebraic derivation of) the usual form. Namely, in our
algebraic formalism the canonical connection from
Proposition~\ref{homog.connection.gen} at least in the stated patch is
 \begin{eqnarray*}
 \omega(Z)& = &\sum Si(Z)\o d i(Z)\t = \cdot(S\tens d)\Delta i(Z)\\
 & = &
 \cdot(S\tens d)\Delta (\delta^{-1}\alpha ) =  \cdot(S\tens
 d){\alpha\tens\alpha+\beta\tens\gamma\over
 \delta\tens\delta+\gamma\tens\beta} \\
 & = &\cdot(S\tens\id)\left({\alpha\tens d\alpha+\beta\tens
 d\gamma\over \delta\tens\delta+\gamma\tens\beta} -
 {(\alpha\tens\alpha+\beta\tens\gamma)\over(\delta\tens\delta +
 \gamma\tens\beta)^2}(\delta\tens d\delta+\gamma\tens d\beta)\right)\\
 &=&\delta d\alpha-\beta d\gamma-\alpha d\delta +\gamma d\beta =
2(\delta d\alpha - \beta d\gamma )
 \end{eqnarray*}
 where we noted that the algebra and calculus are
 commutative and $\delta\alpha-\beta\gamma=1$. The computation is done
 in the algebra of functions on $SU(2)$. The result evidently exists
 globally in this form and can then be cast in the two forms stated.
 The Cartesian co-ordinates $x,y,z$ were given above. Note that the two
trivializations  are connected by a gauge transformation $e^{2\imath\psi} =
{x+\imath y\over x-\imath y} = {-b_+ \over b_-}$, where $\psi$ is the
azimuthal angle. The charge one computation is similar but slightly more
complicated because of the square-roots. \endproof

\note{It is clear that commutative differential structure on
 $P_0(B_0,A)$ is canonical. Explicitly the ideal $M_A$ is generated
by $Z^{-1}+Z - 2$, while $M_P$ is generated by
\begin{eqnarray*}
\delta +\alpha -2 ,\quad \gamma^2 ,\quad \beta\gamma ,\quad \beta^2,
\quad (\alpha-1)\gamma , \quad (\alpha -1)\beta .
 \end{eqnarray*}
 Now
\begin{eqnarray*}
i(Z^{-1}+Z-2) & = & \alpha^{-1}\delta +\delta^{-1}\alpha - 2 \mid
 \cdot \alpha\delta \sim
\delta^2 +\alpha^2 -2\alpha\delta \\
& \sim & \delta^2+\alpha^2 -2 \mid\cdot \delta \sim \alpha + \delta^3
 -2\delta \\
 & \sim & \alpha +\delta(2-\alpha)^2 -2\delta \sim \alpha +4\delta
-4\delta\alpha + \delta\alpha^2 -2\delta \\
 & \sim & 2(\alpha+\delta-2).
 \end{eqnarray*}
 Hence $i(Z^{-1}+Z-2)\in M_P$.}

Now we consider the quantum case. We begin
with the quantum group $SU_q(2)$. It has homogeneous non-commutation
relations:
\[\alpha\beta = q\beta \alpha , \quad \alpha\gamma = q
\gamma\alpha . \quad \alpha \delta = \delta\alpha +
(q-q^{-1})\beta\gamma \]
\[\beta\gamma = \gamma\beta, \quad \beta\delta = q\delta \beta
,\quad \gamma \delta = q \delta \gamma \]
 and a determinant relation
$\alpha\delta-q\beta\gamma=1$. The $*$-structure is
$\alpha^*=\delta$, $\beta^*=-q\gamma$. Of course, these are no longer
functions but abstract elements of the algebra but with analogous
properties. We define
$SO_q(3)$ in the same way as the even elements of this. For $\pi$ and the
resulting coactions we have $\Delta_R$ as
above (unchanged). For generators $b_\pm,b_3$ of $B$ we take the same
expressions as above (unchanged) in terms of $\alpha,\beta$. Their commutation
relations inherited from $SU_q(2)$ are now non-trivial
\[ b_3b_-=(1-q^{-2})b_-+q^{-2} b_-b_3,\quad
b_3b_+=b_+(1-q^{2})+q^{2}b_+b_3\]
\[ b_3^2=b_3+q^{-1}b_-b_+,\quad q^{-2}b_-b_+=q^{2}b_+b_- + (q^{-1} -
q)(b_3-1)\]
and the $*$-algebra structure is $b_\pm^*=-q^{\mp 1} b_\mp$ and
$b_3^*=b_3$. This $B$ is a case of the quantum sphere $S_q^2$ of
Podle\'s \cite{Pod:sph}.

The expressions for $\Phi_i$ are unchanged (but note now that the
order matters).  We proceed for the $SO_q(3)$ case and localise by adjoining
the
same generators as before.

\begin{prop} Let $P=SO_q(3)$, $B=S_q^2$ as above. The localizations
$P_0=SO_q(3)[\delta^{-1}\alpha,\alpha^{-1}\delta]$ over
$B_0=S_q^2[b_3^{-1}]$, and
$P_1=SO_q(3)[\gamma\beta^{-1},\beta\gamma^{-1}]$ over
$B_1=S_q^2[(1-b_3)^{-1}]$ are trivial quantum principal bundles (with universal
differential calculus and
trivializations $\Phi_i$) and paste together in the double localization given
by a trivial bundle
$P_{01}$ over $B_{01}$. We call $P$ over $B$ with these localizations
the {\em quantum monopole bundle}. It is a quantum principal bundle in the
sense of Definition~\ref{principal.bundle}.
\end{prop}
\proof First we construct the nontrivial bundle $P(B,A,\pi)$ using the theory
in Section~5.1.
Since freeness is automatic because $\pi$ is a surjection, we have only to show
the exactness condition. To do this we use Lemma~\ref{homog.bundle} where we
have seen that it suffices to show that $\ker\pi\subset\cdot
(\ker\pi\mid_B\tens P)$. The only
generators for which this  is
non-trivial may be written as follows
\begin{eqnarray}
\beta & = & q^{-1}b_-\delta - (b_3-1)\beta \nonumber\\
\gamma & = & b_+\alpha - q^{-2}(b_3-1)\gamma. \nonumber
\end{eqnarray}
Multiplying on the right by the generators gives the corresponding relations
for elements of $SO_q(3)$. From this it is clear that every
$u\in\ker\pi$ may be expressed as $u=\sum_i b_iv_i$
where $b_i\in\ker\pi\mid_B$ and $v_i\in P$,  hence
$\ker\pi\subset\cdot
(\ker\pi\mid_B\tens P)$. Using Lemma~\ref{homog.bundle} we
deduce that we have a quantum principal bundle (so far with the universal
calculus).
Moreover, we show that the each of the patches shown are trivial bundles and
glue together
by gauge transformations. Firstly, the coaction $\Delta_R$ extends to the
localizations as an
algebra homomorphism, and from this it is clear that $\Phi_i$ are intertwiners.
Since $k[Z,Z^{-1}]$ is
free they extend as algebra maps and are therefore necessarily convolution
invertible. Hence each of the bundles
is trivial from Example~\ref{trivial.bundle}.
Note that this implies that every element of $P_i$ can be written uniquely in
the form
$B_ik[Z,Z^{-1}]$ via the maps $\Phi_i$. This comes out explicitly for $P_0$ as
\[\alpha^2 = b_3 \Phi_0(Z), \quad \alpha\gamma = q b_+\Phi_0(Z),
\quad \gamma^2 = qb_+b_3^{-1}b_+ \Phi_0(Z) \]
\[\beta \delta = q^{-1} b_-\Phi_0(Z^{-1}), \quad \beta^2 = q^{-3}b_3^{-1}b_-^2
\Phi_0(Z^{-1}), \quad \delta^2 = (1-q^{-2} + q^{-2}b_3)\Phi_0(Z^{-1}). \]
{}From the commutation relations
\[ \Phi_{0} (Z)b_3 = (q^{4} b_3  + (1-q^{4}))\Phi_0(Z) \]
\[ \Phi_0(Z) b_- = (q^{4} b_- + q^{2}(1-q^{2}))\Phi_0(Z) \]
\[\Phi_0 (Z) b_+ = (q^{4}b_+ + q^{2} (1-q^{2}))\Phi_0(Z) \]
and linear independence arguments one can verify that all elements of $P_0$ can
similarly be obtained in a unique way. For $P_1\supset B_1$ one has
\[\alpha^2 = q^{2} b_-^2(1-b_3)^{-1}\Phi_1(Z) , \quad \alpha\gamma = -
b_- \Phi_1 (Z) , \quad \gamma^2 = q^{-1} (1-b_3) \Phi_1(Z) \]
\[\beta\delta = - b_+ \Phi_1(Z^{-1}), \quad \beta^2 = q^{-1}(1-b_3)
\Phi_1(Z^{-1}), \quad \delta^2 = b_+(1-b_3)^{-1}b_+ \Phi_1(Z^{-1}) \]
and $\Phi_1$ commutes with $b_3,b_\pm$.

By a similar argument the double localization $P_{01}$ is a
trivial quantum bundle over the double localization $B_{01}$. There are two
trivializations of $P_{01}$,
one is related to $\Phi_0$ while the second to $\Phi_1$. They are both
intertwiners and convolution
invertible. To give the unique decomposition explicitly it
suffices to show that $\gamma \beta^{-1}$ and $\beta \gamma^{-1}$ can
be represented in terms of elements of $B_{01}$ and map $\Phi_{0}$ or
equivalently that $\delta^{-1}\alpha$ and $\alpha^{-1}\delta$ can be
represented in terms of $B_{01}$ and map $\Phi_1$. This comes out as
\[\gamma \beta^{-1} = -q^{2}(1-b_{3})^{-1}b_+b_3^{-1}b_+ \Phi_0(Z)\]
\[\beta\gamma^{-1} = -q^{-2}(1-b_3)^{-1}b_3^{-1}b_-^2\Phi_0(Z^{-1})\]
and
\[\delta^{-1}\alpha = q^{2} b_3^{-1}b_-^2(1-b_3)^{-1}\Phi_1(Z)\]
\[\alpha^{-1}\delta = q^{-2}b_+b_3^{-1}b_+(1-b_3)^{-1}\Phi_1(Z^{-1}).\]

Finally, these two trivializations of $P_{01}$ are equivalent via the gauge
transformation
$$\gamma_{01}(Z)= \Phi_0(Z)\Phi_1(Z)^{-1}$$
(see Proposition~\ref{change.trivialization}), because
\[\gamma_{01}(Z) = -q^{-1}b_+^{-1}b_- = -q^{2}
b_3^{-1}b_-^2(b_3-1)^{-1}  \in B_{01} \]
\[\gamma_{01} (Z^{-1}) = q^{-2} b_+b_3^{-1}b_+(1-b_3)^{-1}\in B_{01}.\]

\endproof

Thus we have a quantum principal bundle (with universal calculus) and a local
trivialization for it.
Next, the argument that the $Ad_R$-covariant function
$i$ must be a combination of $\alpha,\delta$ etc goes through
unchanged and so we can consider $i(Z)=\delta^{-1}\alpha$ as before. In
principle we can proceed formally with
the corresponding canonical connection $\omega$ as above not note that because
the universal differential calculus has no commutation relations betweem
functions and forms on $P$, there is no way to cancel inverses arising in
$\omega$ from $\delta^{-1}$ as was the case in
Proposition~\ref{class.monopole}. One can proceed in the universal case only on
the basis of formal power-series.

Now we come to the details for a non-universal differential calculus, where we
will be able to compute the canonical connection $\omega$ from
Proposition~\ref{homog.connection.local} in closed form. We take for $\Gamma_P$
the left-covariant differential calculus on $SO_q(3)$ inherited from the
left-covariant 3D differential
calculus on $SU_q(2)$ in \cite{Wor:twi}. As $q\to 1$ this tends to the usual
commutative differential calculus in which forms and functions commute. For
convenience we
work in $SU_q(2)$ and afterwards restrict to the relevant subalgebra. The
relevant ideal $M_P \in SU_q(2)$  for generic $q$
 is generated by six elements
\begin{eqnarray}
\delta+q^2\alpha -(1+q^2), \quad \gamma^2, \quad \beta\gamma
\nonumber \\
\beta^2, \quad (\alpha - 1)\gamma , \quad (\alpha - 1)\beta .
\nonumber
\end{eqnarray}

We choose the  basis of the space of the left-invariant 1-forms on $P$
to be
\[\omega^0 = \pi_N\kappa(1\tens\beta),\quad \omega^1=
\pi_N\kappa(1\tens(\alpha-1)),\quad \omega^2 =
-q^{-1}\pi_N\kappa(1\tens\gamma). \]
Explicitly
\begin{eqnarray}
\omega^0 & = & \delta d\beta -q^{-1}\beta d\delta \nonumber \\
\omega^1 & = & \delta d\alpha -q^{-1}\beta d\gamma \\
\omega^2 & = & \gamma d\alpha - q^{-1}\alpha d\gamma \nonumber
\end{eqnarray}
We have the following commutation relations
between $\omega^i$, $i=0,1,2$ and the generators of $SU_q(2)$
\begin{eqnarray}
\omega^0\alpha & = & q^{-1}\alpha\omega^0 , \quad \omega^0\beta =
q\beta\omega^0 \nonumber \\
\omega^1\alpha & = & q^{-2}\alpha\omega^1 , \quad \omega^1\beta =
q^2\beta\omega^1 \label{commut.3d}\\
\omega^2\alpha & = & q^{-1}\alpha\omega^2 , \quad \omega^2\beta =
q\beta\omega^2 \nonumber .
\end{eqnarray}
The remaining relations can be obtained by the replacement
$\alpha\rightarrow\gamma$, $\beta\rightarrow\delta$.
 The relation between
exterior differential $d$ and basic one-forms $\omega^i$  is given
by
\begin{equation}
d\alpha = \alpha\omega^1 -q\beta\omega^2 ,\quad d\beta =
\alpha\omega^0 - q^2\beta\omega^1
\label{def.d}
\end{equation}
and similarly with $\alpha$ replaced by $\gamma$ and $\beta$ replaced
by $\delta$.

\note{Using this data we can construct differential calculus on $SU_q(2)$.
For example the ideal $M_P$ restricted to $SU_q(2)$ is generated by
\[(\delta-\alpha)\gamma ,\quad (\delta-\alpha)\beta ,\quad \gamma^2,
\quad \beta\gamma ,\quad \beta^2 ,\quad \delta^2+q^4\alpha^2 -(1+q^4)
\]}
Projected down to $U(1)$ this gives the ideal $M_A$ generated by
\[ Z^{-1} +q^4 Z -(1+q^4). \]
Obviously this ideal is $Ad_R$-invariant, hence the resulting calculus
is bicovariant as required. The commutation relation in $\Gamma_A$
reads
\begin{equation}
ZdZ = q^{4}dZZ
\end{equation}
One has to check that the 3D calculus fulfills in this way the various
requirements in
Proposition~\ref{homog.bundle.gen} so that we have a quantum homogeneous bundle
in the sense of the general theory developed in earlier sections.

\begin{prop}
Let $P=SO_q(3)$ and $A=k[Z,Z^{-1}]$ with projection $\pi$ be the data as above
for the quantum monopole bundle but equipped now with $M_P$ and the induced
$M_A$ for the 3D differential calculus. Then $P(B,A,\pi,M_P,M_A)$ is a quantum
principal bundle on $B=S_q^2$ in the sense of
Proposition~\ref{homog.bundle.gen}.\label{monopole.bundle.3d}
\end{prop}
\proof By the direct computation one easily finds that
$(id\otimes\pi)Ad_R(M_P)\subset M_P\otimes A$. Explicitly
\begin{eqnarray*}
(id\otimes\pi)Ad_R(\delta +q^2\alpha -(1+q^2)) & = & (\delta +q^2\alpha
-(1+q^2))\tens 1\\
(id\otimes\pi)Ad_R(\gamma^2) & = & \gamma^2 \tens Z^2 \\
(id\otimes\pi)Ad_R(\beta^2) & = & \beta^2 \tens Z^{-2}\\
(id\otimes\pi)Ad_R(\beta\gamma) & = & \beta\gamma \tens 1 \\
(id\otimes\pi)Ad_R((\alpha-1)\gamma)) & = & (\alpha-1)\gamma\tens Z\\
(id\otimes\pi)Ad_R((\alpha-1)\beta)) & = & (\alpha-1)\beta\tens
Z^{-1}.
\end{eqnarray*}
 Moreover $M_A=
\pi(M_P)$ by definition. Hence the hypothesis of Proposition~5.7. is
satisfied and the assertion follows.
\endproof

\begin{prop}\label{prop.connection.3d} The map
\[ \omega(a)=\sum Si(a)\o d i(a)\t\]
is a connection 1-form on the quantum monopole bundle for the 3D calculus
in Proposition~\ref{monopole.bundle.3d}. In terms of one forms
$\omega^i$ it can be written explicitly as
 \begin{equation}
\omega(f(Z)) = [2]_{q^{-2}} D_{q^{-4}}f(Z)\mid_{Z=1}\omega^1,
\label{omega.charge2}
\end{equation}
where we used the by now standard notation $[n]_x=\frac{x^n-1}{x-1}$,
$f(Z)$ represents a general element of $A$
understood as a Laurent
series in variable $Z$, and $D_x$ is the Jackson's derivative labelled
by $x$, i.e.
\begin{equation}
D_x(f(Z)) = {{(f(xZ) -f(Z))}\over{(x-1)Z}}.
\end{equation}
\end{prop}
\proof We show that $i(M_A)\subset M_P$. From Proposition~5.8 we then
deduce that $\omega$ is a connection 1-form. First we notice that
$\delta^2 +q^4\alpha^2 - (1+q^4)\in M_P$. Next, applying $i$ to the
generator of $M_A$ we find
\begin{eqnarray*}
i(Z^{-1} +q^4Z - (1+q^4)) & = & \alpha^{-1}\delta
+q^4\delta^{-1}\alpha - (1+q^4) \\
& = &  \alpha^{-1}\alpha\delta\delta  - q\alpha^{-1}\beta\gamma\delta
+ q^4\delta^{-1}\delta\alpha\alpha - q^3\delta^{-1}\beta\gamma\alpha -
(1+q^4) \\
& = & \delta^2 +q^4\alpha^2 - (1+q^4) -
\beta\gamma(q^{-1}\alpha^{-1}\delta + q^5\delta^{-1}\alpha) \in M_P.
\end{eqnarray*}
According to
Proposition~5.8. $\omega$ is a connection 1-form and hence there is a
map $\sigma_N:P\otimes\ker\eps/M_A\rightarrow\Gamma_P$ such that
\begin{equation}
\omega (a) = \sigma_N (1\otimes \pi_A (a - \eps(a))).
\end{equation}
Using definition of the ideal $M_A$ it is easy to compute
\begin{equation}
\pi_A (f(Z)-f(1)) = D_{q^{-4}}f(Z)\mid_{Z=1}\pi_A(Z-1).
\end{equation}
Hence
\begin{equation}
\omega (f(Z)) =  D_{q^{-4}}f(Z)\mid_{Z=1}\omega(Z).
\end{equation}
Now it remains to compute $\omega(Z)$ explicitly. First we notice that
\begin{equation}
\omega^1 =  ([2]_{q^{-2}})^{-1} \pi_N\kappa^{-1}(1\tens(\delta^{-1}\alpha-1)).
\label{omegi}
\end{equation}
This follows from the fact that
\[0\sim\delta+q\alpha-q\mu\sim\delta\alpha+q^2\alpha^2-(1+q^2)
\alpha\sim 1+q^2\alpha^2-(1+q^2)\alpha. \]
and that
\[\delta^{-1}\alpha = \delta^{-1}\delta\alpha^2 -
q^{-3}\beta\gamma\delta^{-1}\alpha \sim \alpha^2 . \]
The symbol $\sim$ means that we identify two elements of $\ker\eps_P$
if they differ by an element in $M_P$, and we used that
\[\alpha\delta = 1+q\beta\gamma \sim 1 \sim \delta\alpha . \]
On the other hand we know that $\omega$ is given by
(\ref{omega.kappa}). For $a=Z$ we find
\begin{eqnarray}
\omega(Z) & = & \pi_N\kappa(1\otimes i(Z-1)) =
\pi_N\kappa(1\otimes (\delta^{-1}\alpha -1)) \nonumber \\
& = & [2]_{q^{-2}}\omega^1 = [2]_{q^{-2}} (\delta d\alpha -
q^{-1}\beta d \gamma).\nonumber
\end{eqnarray}
Hence finally,
\[\omega(f(Z)) =  [2]_{q^{-2}} D_{q^{-4}}f(Z)\mid_{Z=1}\omega^1\]
as stated.
\endproof

We observe that $\omega$ admits the following local representation
(compare Proposition~\ref{class.monopole})
\[ \omega(Z)=\cases{\beta_0(Z)+\Phi_0^{-1}(Z)d\Phi_0(Z),\quad
\beta_0(Z)=qb_3^{-1}(q^2b_+d b_- - q^{-2}b_- d b_+ - \lambda d
b_3)&\cr
\beta_1(Z)+\Phi_1^{-1}(Z)d\Phi_1(Z),\quad \beta_1(Z)=q(b_3-1)^{-1}(q^2b_+d b_-
- q^{-2}b_-d b_+ - \lambda d
b_3)&}\]
where $\lambda = q-q^{-1}$.

This completes our treatment of the charge two monopole. To conclude we discuss
the situation for the connection 1-form corresponding to the
charge one monopole as discussed in the classical situation. Firstly, there is
no problem to
construct the bundle $P(B,A,\pi,M_P,M_A)$ with $P=SU_q(2)$, $A= U(1)$,
$B=S_q^2$,
$\pi$, $M_P$ and $M_A$ as before. We have already done the relevant
computations. On the other hand, to define local trivializations of
$P(B,A,\pi,M_P,M_A)$ and eventually the map $i$ one has to formally adjoin the
square roots $\sqrt{\delta^{-1}\alpha}$, $\sqrt{\alpha^{-1}\delta}$ to
$P$. Assuming this, one can define the map $i:A\rightarrow P_0$ by
\[i(Z^\h) = \sqrt{\delta^{-1}\alpha}, \quad i(Z^{-\h})=
\sqrt{\alpha^{-1}\delta}\]
and argue that $i(M_A)\subset M_P$. We have
\begin{eqnarray*}
&&i(Z^{-\h} + q^2Z^\h - (1+q^2) ) = \sqrt{\alpha^{-1}\delta}
+q^2\sqrt{\delta^{-1}\alpha} - (1+q^2) \\
&& = \sqrt{\alpha^{-1}\alpha\delta^2-q\alpha^{-1}\beta\gamma\delta}
+q^2\sqrt{\delta^{-1}\delta \alpha^2 -
q^{-1}\delta^{-1}\beta\gamma\alpha} -(1+q^2) \\
&& =  \left(\sqrt{1-q^{-1}\beta\gamma(\delta\alpha)^{-1}}\right)\delta +
q^2\left(\sqrt{1-q\beta\gamma(\alpha\delta)^{-1}}\right)\alpha - (1+q^2) \\
&& = \delta + q^2\alpha - (1+q^2) - \sum_{n=1}^{\infty}
c_n(\beta\gamma)^n(q^{-n}(\delta\alpha)^{-n}\delta +
q^{n+2}(\alpha\delta)^{-n}\alpha)\in M_P
\end{eqnarray*}
where $c_n$ are coefficients in the power series expansion
\[\sqrt{1-x} = 1 - \sum_{n=1}^{\infty}c_nx^n.\]
For this reason the computation of the charge one monopole is formal.

Proceeding formally we next apply Proposition~\ref{homog.connection.local} and
deduce that there is a canonical connection in the bundle
$P(B,A,\pi,M_P,M_A)$. We can compute its connection 1-form explicitly,
using the same methods as before. First we notice that
\[\pi_A(f(Z^\h)-1)=D_{q^{-2}}f(Z^\h)\mid_{Z^\h=1}\pi_A(Z^\h-1).\]
Hence from the definition of the the connection 1-form we deduce that
\[\omega(f(Z^\h)) = D_{q^{-2}}f(Z^\h)\mid_{Z^\h=1}\omega(Z^\h).\]
Finally we notice that
\[\sqrt{\delta^{-1}\alpha} = \alpha -
\sum_{n=1}^{\infty}c_nq^n(\beta\gamma)^n(\alpha\delta)^{-n}\alpha \sim
\alpha \]
so that
\[\omega(Z^\h) = \pi_{N_P}\kappa(1\otimes i(Z^\h-1)) =
\pi_{N_P}\kappa(1\otimes(\alpha-1)) = \omega^1.\]
Therefore
\begin{equation}
\omega(f(Z^\h))=D_{q^{-2}}f(Z^\h)\mid_{Z^\h=1}\omega^1.
\label{omega.charge1}
\end{equation}
Comparing this result with  (\ref{omega.charge2}) we see that the
quantum integer $[2]_{q^{-2}}$ has a natural interpretation as the
q-monopole charge. Note that the power appearing in the
expression  for $i$ corresponds to the winding number in the
classical situation, which is the topological interpetation of the monopole
charge. A corresponding picture in the quantum case, as well as the
construction of
higher monopole charges, are interesting directions for further work.

In addition, it is hoped to give some concrete applications of this
construction
along lines sketched in the introduction. For example we note that non-trivial
superselection sectors for quantum mechanics on
$S_q^2$ have recently been detected in \cite{Egu:mec} and it would be
interesting to try to relate them to
our quantum monopole bundle. Our constructions are not tied to this example and
with suitable projections and inclusions can be used for other quantum groups
and their canonical connections just as well. For example, a natural next goal
would the construction of a q-deformed instanton based on these techniques. The
first problems for this are quantum-group theoretical (one needs the analogues
of usual groups and their inclusions), and will be attempted elsewhere.

\appendix
\section{Appendix: quantum associated vector bundles}

In this appendix we develop the non-commutative analogue of the following
classical theory. This is needed to tie our theory in Section~4.1 to the local
picture in Section~3.

Let $P(M,G)$ be a usual principal bundle and let $V$ be a vector space and
$\rho$ a representation of $G$ on
$V$. Any $V$-valued form $\phi$ on $P$ such that
\begin{equation}
(R_{a}^{*} \phi )(X) \equiv \phi ((R_{a})_{*} X) = \rho (a^{-1})
\phi (x)
\label{pseudotensorial.class}
\end{equation}
is called a pseudotensorial form on $P$. A pseudotensorial form $\phi$ on
$P$  is said to be tensorial if it vanishes on horizontal vectors (it
corresponds to the section of a bundle associated to $P$). If $\phi$
is a tensorial form then we can  define covariant derivative on $\phi$
by
$$D \phi = d\phi \circ (1 - \widetilde{\ } \circ \omega)$$
i.e.
$$D_{X} \phi = i_{X} d \phi - i_{\widetilde{\  } \omega (X)} d\phi = i_{X}
d\phi + \rho (\omega (X)) \phi$$
where $\omega $ is a connection 1-form, $X$ is a vector field and $i$ denotes
interior product (evaluation).

For any principal bundle $P(M,G)$ and vector space $V$ on which $G$
acts, we can define the associated vector bundle
$E(M,V,G)$ with fibre $V$. Let $\rho$ be the representation of $G$ on $V$
and define the equivalence relation $\sim$ on $P \times V$ given by $(u,v)
\sim (ua, \rho (a^{-1})v)$. The total space $E$ of the bundle
$E(M,V,G)$ associated to $P$ is the quotient of $P \times V$ by the
relation $\sim$. In local coordinates:
$$E \cong (M \times G)_{G} \times V \cong M \times (G \times_{G} V)
\cong M \times V.$$

We now develop the quantum picture, working for simplicity in the case of
universal differential
calculus. Let $P(B,A)$ be a quantum principal bundle as defined in
Definition~\ref{principal.bundle} and let $\Pi$ be a connection in the
principal bundle $P$. We
define
horizontal n-forms on $P$ to be elements of the set $\Omega ^nP_{hor} =
P j(\Gamma_{B}) P j(\Gamma_B) P \cdots P j(\Gamma_B) P$  (n times). The space
of all horizontal forms will be denoted by $\Omega P_{hor}$. We say
that a form $\alpha \in \Omega P$ is {\em
strongly horizontal} if $\alpha \in j(\Omega B)P$. We write $\Omega
P_{shor} \equiv j(\Omega
B)P$. Note that $\Omega P_{shor} \subset \Omega P_{hor}$.

\begin{prop}
If the bundle $P(B,A)$ has a
connection $\Pi$, then the map
\begin{equation}
h(u_0du_1 \cdots du_n) = u_0 (id- \Pi)(du_1)(id- \Pi)(du_2) \cdots
(id-\Pi)(du_n)
\label{proj.horizontal}
\end{equation}
where $u_0, \ldots , u_n \in P$, is a linear projection of $\Omega P$
onto $\Omega P_{hor}$. Moreover,
\begin{equation}
\Delta_{R} h = ( h \otimes id) \Delta_R.
\label{proj.horizontal.1}
\end{equation}
\end{prop}
\proof
\note{To prove that the map $h$ is well defined, we assume that
\begin{equation}
\sum_{i} u_{0i}du_{1i} \cdots du_{ni} = 0
\label{horizontal.proj.proof.1}
\end{equation}
for some $u_{0i}, \ldots , u_{ni} \in P$. Let us observe that, without
loss of generality we can assume that $u_{1i}, \ldots ,u_{ni}$ are in
the non-unital part of $P$ and that the set $\{ du_{n1}, du_{n2},
\ldots \}$ is linearly independent. The set $\{u_{n1},
u_{n2}, \ldots \}$ is linearly independent. Using the explicit definition
of the universal differential we see that equation
(\ref{horizontal.proj.proof.1}) implies that
\begin{equation}
\sum_{i} u_{0i} \otimes_k u_{1i} \otimes_k \cdots
\otimes_k u_{ni} = 0
\end{equation}
i.e. $u_{0i}\otimes_k u_{1i} \otimes_k \cdots \otimes_k  u_{n-1i} = 0$
for every $i$.
This in turn implies that at least one of the factors in the last tensor
product must vanish, therefore the
right-hand side of (\ref{proj.horizontal}) also vanishes. Thus the map
$h$ is well-defined.}
It is easy to see that the map $h$ is well-defined as stated. It  is a
projection because every $(id-\Pi)$ is a projection and
$h(\Omega ^{n}P) = \Omega^nP_{hor}$ as $(id-\Pi)(\Gamma_{P}) =
P\Gamma_BP$. Finally  equation (\ref{proj.horizontal.1}) can be checked
directly as
\begin{eqnarray*}
\Delta_R h(u_{0} du_1 \cdots du_n) & = & \Delta_R (u_0 (id-\Pi)(du_1)
\cdots (id- \Pi)(du_n))\\
 & = &\Delta_R (u_0) \Delta_R ((id-\Pi)(du_1))
\cdots \Delta_R ((id - \Pi)(du_n)) \\
& = & u_0\bo (id -\Pi)du_1\bo \cdots (id - \Pi)du_n\bo \otimes u_0\bt
u_1\bt\cdots u_n\bt \\
& = & (h \otimes id)\Delta_R (u_0du_1 \cdots du_n).
\end{eqnarray*}
Here the third equality uses covariance of the universal envelope $\Omega
P$ and invariance of the connection $\Pi$ (see (\ref{covariance}) and
(\ref{inv.conn})).
\endproof

% In the non-universal case, we take as an additional assumption
% that $h$ can be defined  on $\Omega (P)$  in  such a way that
% (\ref{proj.horizontal.1}) is obeyed.

Let $(V, \rho_{R})$ be a right $A^{\rm op}$-comodule algebra, and let
$\phi : V \rightarrow \Omega P$ be a linear map. We say that $\phi$
is a {\em pseudotensorial form on $P$} if
\begin{equation}
\Delta_{R} \phi = (\phi \otimes id ) \rho_{R}.
\label{pseudotensorial}
\end{equation}
A map $\phi : V \rightarrow \Omega P$ is
called a {\em tensorial form on $P$} ({\em strongly tensorial form on
$P$}) if it is pseudotensorial and for
any $v \in V$, $\phi (v)$ is horizontal (strongly horizontal resp.)
(compare eq. (\ref{pseudotensorial.class})).

\begin{lemma}
Let $\phi : V \rightarrow \Omega P$ be a tensorial form on
a quantum principal bundle $P(B,A)$ with connection $\Pi$. Then
$d \phi: V \rightarrow \Omega P$ is pseudotensorial.
\end{lemma}
\proof
To prove the lemma we need only note that
\begin{eqnarray*}
\Delta_{R} (d\phi) = (d \otimes id) \Delta_{R} \phi = (d \otimes id)
(\phi \otimes id) \rho_{R} = (d\phi \otimes id) \rho_{R}.
\end{eqnarray*}
\endproof

The map
\begin{equation}
D = hd
\label{cov.der.def.1}
\end{equation}
is called the {\em exterior covariant derivative in $P$}. Here $D$ sends
tensorial forms into  tensorial forms (since the projection $\Pi$ is right
invariant).

We can now define the notion of a quantum vector bundle associated to
a quantum principal bundle $P(B,A)$.

\begin{df}
Let $P(B,A)$ be a quantum principal bundle and let $V$ be a right
$A^{\rm op}$-comodule algebra with coaction $\rho_{R} : V \rightarrow V
\otimes A$. The space $P \otimes V$ is naturally endowed with a right
$A$-comodule structure  $\Delta_{E} : P \otimes V
\rightarrow P \otimes V \otimes A$ given by
$$\Delta_{E} (u \otimes v) = \sum u^{(\overline 1)} \otimes
v^{(\overline 1)} \otimes u^{(\overline 2)}v^{(\overline 2)}$$
for any $u \in P$ and $v \in V$. We say that the space
$$E = (P \otimes V) ^{A} = \{ u\otimes v \in P \otimes V : \Delta_{E}
(u \otimes v) = u \otimes v \otimes 1 \}$$
is a  {\em quantum vector bundle associated to $P$} over $B$ with
structure group $A$ and standard fibre $V$. We denote it by $E =
E(B,V,A)$.
\end{df}

\begin{lemma}

\begin{enumerate}
\item $E$ is a subalgebra of $P \otimes V$.
\item $B$ is a subalgebra of $E$.
\end{enumerate}
\end{lemma}
\proof To prove the first assertion let us take $u_{1} \otimes
v_{1}, \; u_{2} \otimes v_{2} \in E$. Then we have
\begin{eqnarray*}
\Delta_{E} (u_{1} u_{2} \otimes v_{1} v_{2}) &=& \sum
u_{1}^{(\overline 1)}u_{2} ^{(\overline 1)} \otimes v_{1}^{(\overline
1)} v_{2}^{(\overline 1)} \otimes u_{1}^{(\overline
2)}u_{2}^{(\overline 2)}v_{2}^{(\overline 2)}v_{1}^{(\overline 2)} \\
& = & \sum (u_{1}^{(\overline 1)} \otimes v_{1}^{(\overline 1)}
\otimes u_{1}^{(\overline 2)}) (u_{2}^{(\overline 1)} \otimes
v_{2}^{(\overline 1)} \otimes u_{2}^{(\overline 2)}v_{2}^{(\overline
2)}) (1 \otimes 1 \otimes v_{1}^{(\overline 2)})\\
&=& \sum (u_{1}^{(\overline 1)} \otimes v_{1}^{(\overline 1)} \otimes
u_{1}^{(\overline 2)}) (u_{2} \otimes v_{2} \otimes 1)(1 \otimes 1
\otimes v_{1}^{(\overline 2)})\\
&=& u_{1}u_{2} \otimes v_{1}v_{2} \otimes 1
\end{eqnarray*}
Hence $(u_{1} \otimes v_{1})(u_{2} \otimes v_{2}) \in E$, and
$E$ is a subalgebra of $P \otimes V$ as stated. To prove the second
statement of the lemma let us observe that there is a map $j_{E}: B
\hookrightarrow P \otimes V$ defined by $j_{E}(b) = b \otimes 1_{V}$ for any $b
\in B$ and $j_{E}(b) \in E$ since
$$\Delta_{E} j_{E} (b) = \Delta_{E} (b \otimes 1_{V}) = b \otimes
1_{V} \otimes 1_{A} = j_{E}(b) \otimes 1.$$
This proves the lemma. \endproof

Let $E(B,V,A)$ be a quantum vector bundle associated to $P(B,A)$. We
say that a map $s: E \rightarrow B$ is a {\em cross-section} of E if:
\begin{equation}
s \circ j_{E} = id
\label{cross-section}
\end{equation}

\begin{prop}
Let $\phi : V \rightarrow P$ be a pseudotensorial 0-form on $P$ such
that $\phi (1_{V}) = 1_{P}$. Then the map $s: E \rightarrow B$ given
by
\begin{equation}
s = \cdot (id_{P} \otimes \phi)\mid_{E}
\end{equation}
is a cross-section of $E$.
\end{prop}
\proof First we show that $s$ takes its values in $B$.
Take $u \otimes v \in E$, where $u\in P$, $v \in V$. By the definition
of $E$,
$$\Delta_{E} (u \otimes v) = u \otimes v \otimes 1.$$
Hence
\begin{eqnarray*}
\Delta_{R}s(u \otimes v) & = & \Delta_{R} (\cdot (id_{P} \otimes
\phi)(u \otimes v)) = \sum u^{(\overline 1)}\phi (v^{(\overline 1)})
\otimes u^{(\overline 2)}v^{(\overline 2)}\\
& = & (\cdot \otimes id_{A})(id_{P} \otimes \phi \otimes id_{A})
\Delta_{E} (u \otimes v) = u \phi (v) \otimes 1.
\end{eqnarray*}
Thus $s(x) \in B$ for any $x \in E$. Next we show that $s$ is
a cross-section of $E$. We have
$$s \circ j_{E} (b) = s(j(b) \otimes 1) = j(b) \phi (1) = j(b) =b$$
for any $b \in B$. The last equality is a consequence of the fact that
the inclusion $j$ is just the identity on $B$. \endproof

Let us assume now that we have a trivial bundle $ P(B,A,\Phi)$ as defined in
Example~\ref{def.principal.trivial}
and moreover that our Hopf algebra
$A$ has bijective antipode. Then the map $\Phi: A \hookrightarrow P$
induces naturally a map $\Phi_{E} : V
\hookrightarrow E$, given by
$$\Phi_{E} (v) = \sum \Phi (S^{-1} v^{(\overline 2)}) \otimes
v^{(\overline 1)}$$
for any $v \in V$. This map obiously takes its values in $P\otimes
V$. We want to show that $\Phi_{E}(v) \in E$ for any $v \in V$. We
have
\begin{eqnarray*}
\Delta_{E} \Phi_{E} (v) & = & \sum \Delta_{E} (\Phi (S^{-1}
v^{(\overline 2)}) \otimes v^{(\overline 1)}) \\
& = & \sum \Phi {(S^{-1} v^{(\overline 2)})}^{(\overline 1)} \otimes
v^{(\overline 1) (\overline 1)} \otimes \Phi {(S^{-1} v^{(\overline
2)})}^{(\overline 2)}  v^{(\overline 1)(\overline 2)}
\end{eqnarray*}
but since $\Phi$ is an intertwiner of $\Delta_{R}$ and $\Delta$, we
obtain
\begin{eqnarray*}
\Delta_{E}\Phi_{E} (v) & = &  \sum \Phi (S^{-1} {v^{(\overline
2)}}_{(3)}) \otimes v^{(\overline 1)} \otimes (S^{-1}{v^{(\overline
2)}}_{(2)}) {v^{(\overline 2)}}_{(1)} \\
& = & \sum \Phi (S^{-1} v^{(\overline 2)}) \otimes v^{(\overline 1)}
\otimes 1 = \Phi_{E} (v) \otimes 1.
\end{eqnarray*}
Hence $\Phi_{E}(v) \in E$ for any $v \in V$. Notice also that
 $\Phi_{E} (1_{V}) = 1_{E}$ because of the second of the equations
(\ref{trivialization}).

Moreover, using an analogous proof to that in  Example~\ref{trivial.bundle} we
see that the map
\eqn{trivial.E}{ \theta:B \otimes V \rightarrow E,\qquad \theta(b \otimes
v)=j_{E}(b)\Phi_{E}(v)}
is an isomorphism of vector spaces.  Explicitly, the required inverse map is
\eqn{trivial.Einv}{ \theta^{-1}(u\tens v)=\sum u\bo\Phi^{-1}(u\bt)\tens v
=\sum u\Phi^{-1}(S^{-1}v\bt)\tens v\bo}
where the second form follows since $u\tens v$ lies in $E=(P\tens V)^A$.
Accordingly, we call $E$ in this case a trivial associated
vector bundle and $\Phi_E$ its trivialzation.

\begin{prop}
Let $E(B,V,A)$ be the trivial vector bundle associated to a trivial quantum
principal bundle $P(B,A,\Phi)$ as explained. If $s: E \rightarrow B$ is  a
cross-section of $E$ then
the map $\phi : V \rightarrow P$
\begin{equation}
\phi (v) = \sum j \circ s \circ \Phi_{E} (v^{(\overline 1)}) \Phi
(v^{(\overline 2)})
\label{phi.form.1}
\end{equation}
is a tensorial 0-form on $P$.
\label{section.tensorial.prop.1}

\end{prop}
\proof We need to show that $\phi : V \rightarrow P$ defined by
(\ref{phi.form.1})
is an intertwiner between the coaction $\Delta_{R}$ and the corepresentation
$\rho_{R}: V \rightarrow V \otimes A$. Using (\ref{trivialization})
we obtain
\begin{eqnarray*}
\Delta_{R} \phi (v) & = & \sum (j \circ s \circ \Phi_{E}
(v^{(\overline 1)}) \otimes 1) (\Phi ({v^{(\overline 2)}}_{(1)})
\otimes {v^{(\overline 2)}}_{(2)}) \\
& = & \sum j \circ s \circ \Phi_{E} (v^{(\overline 1)}) \Phi
({v^{(\overline 2)}}_{(1)}) \otimes {v^{(\overline 2)}}_{(2)} = \sum \phi
(v^{(\overline 1)}) \otimes v^{(\overline 2)}.
\end{eqnarray*}
\endproof

We now look at the description of quantum bundles in local
coordinates. For this we restrict ourselves from now to trivial bundles.
We would like to show how the general theory developed above reduces to
the theory described in Section~3 (when the bundles considered were
all trivial). The gauge transformations encountered there will appear now
as transformations of the local description.

\begin{prop}\label{section.tensorial.prop.2}
Let $P(B,A,\Phi)$ be a trivial quantum principal bundle. Let $(V,\rho_{R})$ be
a right
$A^{\rm op}$-comodule algebra and let $\sigma :V \rightarrow \Omega B$
be any linear map. Then the map $\phi : V \rightarrow \Omega P$ given
by
\begin{equation}
\phi (v)= \sum (j \circ \sigma) (v^{(\overline 1)}) \Phi
(v^{(\overline 2)})
\label{section.tensorial.1}
\end{equation}
is a pseudotensorial form on $P$. Conversely, if $\phi : V \rightarrow
\Omega P$ is a strongly tensorial form on $P$ then
$$\sigma (v) = \sum \phi (v^{(\overline 1)}) \Phi^{-1}(v^{(\overline
2)})$$
defines a linear map $\sigma: V \rightarrow \Omega B$
which reproduces $\phi$ according to (\ref{section.tensorial.1}).
\end{prop}
\proof To prove the first assertion we have to check that $\phi$ as
defined is an intertwiner. We have
\begin{eqnarray*}
\Delta_{R} \phi (v) & = & \sum \Delta_{R}(j \circ \sigma (v^{(\overline
1)}) \Delta_{R} \Phi (v^{(\overline 2)})) = \sum (j \circ \sigma
(v^{(\overline 1)}) \otimes 1 )(\Phi ({v^{( \overline 2)}}_{(1)})
\otimes {v^{(\overline 2)}}_{(2)}) \\
& = & \sum j \circ \sigma (v^{(\overline 1)})\Phi ({v^{(\overline
2)}}_{(1)}) \otimes {v^{(\overline 2)}}_{(2)} = (\phi \otimes id)
\rho_{R}.
\end{eqnarray*}
Conversely, we need to prove
that $\sigma (v) \in \Omega B$
for any $v \in V$. But $\sigma (v)$ is strongly horizontal since $\phi
(v)$ is strongly horizontal, i.e. $\sigma (v) \in j(\Omega B) P$.
Moreover,
\begin{eqnarray*}
\Delta_{R} \sigma (v) & = & \sum (\phi (v^{(\overline 1)}) \otimes
{v^{(\overline 2)}}_{(1)})(\Phi^{-1}({v^{(\overline 2)}}_{(3)}) \otimes
S{v^{(\overline 2)}}_{(2)}) \\
& = & \sum \phi (v^{(\overline 1)}) \Phi ^{-1} (v^{(\overline 2)})
\otimes 1.
\end{eqnarray*}
Therefore $\sigma (v)$ is invariant, and since $\Omega B$ contains
any  invariant subset of $j(\Omega B) P$, we conclude that $\sigma (v)
\in \Omega B$. Finally, using the fact that $j$ is the identity on
$\Omega B$ we obtain
$$\sum j\circ\sigma(v\bo)\Phi(v\bt) = \sum \phi(v^{(\overline 1)}) \Phi^{-1}
({v^{(\overline
2)}}_{(1)}) \Phi ({v^{(\overline 2)}}_{(2)}) = \phi(v). $$
\endproof

Composing Proposition~\ref{section.tensorial.prop.1} with
Proposition~\ref{section.tensorial.prop.2}
we obtain:

\begin{cor}
Let $E(B,V,A)$ be the trivial quantum vector bundle associated to a trivial
quantum principal
bundle $P(B,A,\Phi)$. Then any map $\sigma: V \rightarrow B$ such that
$\sigma (1_{V}) = 1_{B}$ induces a cross-section $s: E \rightarrow B$.
Conversely any cross-section $s$ of $E$ induces a map $\sigma: V
\rightarrow B$.
\label{cor.section}
\end{cor}
{\bf Proof} This follows from the above, but a direct proof is also
instructive. Namely, we consider
the trivialization $\Phi_{E} : V \rightarrow E$ and use the isomorphism
$\theta$ in (\ref{trivial.E}). It is evident that
$\theta^{-1}(j_E(b))=b\tens 1$.
Let $\sigma : V \rightarrow B$ be any map such that $\sigma (1_{V})
=1_{B}$ and let $s = \cdot (id \otimes \sigma) \circ \theta^{-1}$.
Obviously $s: E \rightarrow B$. Moreover
$$s \circ j_{E}(b) = \cdot (id \otimes \sigma) \theta^{-1}(j_{E}(b)) =
\cdot (id \otimes \sigma)(b \otimes 1) =b.$$
Thus $s$ is a section on $E$. Conversely if $s$ is any section of $E$
then we define $\sigma = s \circ \Phi_{E}$. \endproof

Now we consider gauge transformations as defined by a change in trivialization.
Such a gauge transformation $\gamma$
also changes the coordinates in the quantum vector bundle $E(B,V,A)$
associated to $P$, inducing  a transformation of sections of $E$,
where the latter are identified with maps $\sigma : V \rightarrow B$ by
Corollary~\ref{cor.section}.

\begin{prop}\label{change.sigma}
Let $P(B,A,\Phi)$ be a trivial quantum principal bundle and  $(V, \rho
_{R})$ a right $A^{\rm op}$-comodule algebra. Let $\sigma : V
\rightarrow B$ be a map
defining a tensorial 0-form $\phi$ by
Proposition~\ref{section.tensorial.prop.2}, and let
$\gamma : A \rightarrow B$ be a gauge transformation. Then
the transformation $\sigma \mapsto \sigma^\gamma = \sigma * \gamma$
for a fixed trivialization $\Phi$ induces a gauge transformation $\phi
\mapsto \phi^\gamma$.
This can also be understood as a transformation of $\Phi$ with fixed $\sigma$,
\[ \phi^\gamma = j(\sigma) \Phi^\gamma. \]

Conversely if $\phi$ is a fixed tensorial 0-form on $P$ and the map
$\sigma : V
\rightarrow B$ is obtained from $\phi$ by
Proposition~\ref{section.tensorial.prop.2}, then a gauge
transformation of the trivialization $\Phi \mapsto \Phi ^\gamma$ induces
a transformation of the local description
\[ \sigma \mapsto \sigma^{\gamma^{-1}} = \sigma * \gamma ^{-1}. \]
\end{prop}
{\bf Proof} This is  by direct computation using the fact that $j$ is
an algebra map. The first statement is
\begin{eqnarray*}
\phi^\gamma \equiv j(\sigma^\gamma) * \Phi = j(\sigma * \gamma) * \Phi
= j(\sigma ) *\Phi^\gamma.
\end{eqnarray*}
For the converse let us observe that $({\Phi ^\gamma})^{-1} = \Phi
* j(\gamma^{-1})$. Then
\begin{eqnarray*}
\sigma^{\gamma^{-1}}= \phi * ({\Phi ^\gamma})^{-1} = \phi * \Phi^{-1} *
\gamma^{-1}
= \sigma * \gamma^{-1}
\end{eqnarray*}
because $j$ is the identity map on $B$. \endproof

The first part of the proposition represents the active point of view
on gauge transformations of principal bundles, while the
second represents the passive point of view. From the latter point of
view, gauge
transformations are automorphisms of the bundle $P$.

Let us note that the transformation law for a map $\sigma$ (from the
active point of view),  is exactly
the same as that given in equation (\ref{trivial.sigma.right}) in Section~3.

Let us finally compute an explicit formula for the covariant derivative
in the case of trivial bundles (to compare it with
(\ref{trivial.nabla.1}) and
(\ref{trivial.nabla.2})). Thanks to
Proposition~\ref{section.tensorial.prop.2}
we know the form of any strongly
tensorial form on $P$. We can define a linear operator  $\nabla$ in
the space of maps $\sigma : V \rightarrow \Omega B$ by means of
\begin{equation}
D \phi = j(\nabla \sigma) * \Phi
\end{equation}
where $\phi$ is a strongly tensorial form and $\sigma$ is a map
decomposing $\phi$ according to (\ref{section.tensorial.1}). We have:
\begin{lemma}
Let $P(B,A,\Phi)$ be a trivial quantum principal bundle with differential
structure given by $\Omega P$. Let $\omega$ given by
(\ref{trivial.connection.beta}) define a connection in $P$. Then for any
$\sigma : V \rightarrow \Omega^n B$ we have
\begin{equation}
\nabla \sigma = d \sigma - (-1)^n \sigma * \beta.
\end{equation}
\end{lemma}
\proof
Using the definition of the covariant derivative $D$ in equation
(\ref{cov.der.def.1}) we compute
\begin{eqnarray*}
D(\sigma * \Phi) & = & h(d\sigma * \Phi + (-1)^n \sigma * d\Phi) =
d\sigma * \Phi + (-1)^n \sigma * d\Phi - (-1)^n \sigma * \Pi_\omega
(d\Phi) \\
& = & d\sigma * \Phi + (-1)^n\sigma *d\Phi - (-1)^n \sigma * \beta *
\Phi - (-1)^n \sigma * d \Phi \\
& = & (d\sigma - (-1)^n \sigma * \beta)
*\Phi
\end{eqnarray*}
as required.
\endproof
\vspace{12pt}

Thus we have obtained from the abstract theory the local picture quoted at the
end of Section~3, at least for the universal calculus.

\section{Appendix: quantum matrix case of the local picture}

Here we collect some results concerning trivial quantum vector bundles
in the case when the structure quantum group is of matrix type. Let
$A$ be such a quantum group
generated by the matrix $\vect= ({t^i}_j)_{i,j=1}^{n}$ obeying some
commutation relations (see \cite{FRT:lie}). There is a natural
comultiplication
in $A$ given by matrix multiplication (we assume summation over
repeated indices), namely $\Delta {t^i}_j = {t^i}_k
\otimes {t^k}_j$. The counit is $\epsilon {t^i}_j = {\delta ^i}_j$.
For example, we can begin with the
matrix bialgebra  $A(R)$ defined by the solution
$R$ of Yang-Baxter equation:
\[R_{12} R_{13}R_{23} = R_{23}R_{13}R_{12}.\]
Here $R \in End (k^{n} \otimes k^{n})$ and $R_{12} = R
\otimes I$ etc. where $k$ is our field (such as $k={\Bbb C})$. The
commutation relations of
$A(R)$ are given by the equation
\[R\vect_1\vect_2 = \vect_2\vect_1R\]
and in nice cases lead to Hopf algebras $A$ after quotienting $A(R)$
by suitable `determinant-type' relations.

We can also obtain examples of suitable fibers from the same matrix $R$ by
setting  $V = Z(R)$, the Zamolodchikov  algebra generated by the set
$\vecv = (v^i)_{i=1}^n$, obeying the relations and left $A(R)$-coaction
\[Rv_1v_2 = \lambda v_2v_1,\quad \rho_L v^i = {t^i}_j \otimes v^j. \]
where $\lambda \in k^*$ is a parameter. One can easily check
that $Z(R)$ is indeed a left $A(R)$-comodule algebra with coaction $\rho_L$.
It was explicitly done in \cite[Sec. 6.3.2]{Ma:qua} in these
conventions. We suppose this quotients also to a coaction of $A$.

If $B$ is any algebra with unit we define the trivial left quantum
vector bundle
$E(B,Z(R),A)$  as in Section~3 and we
keep the formalism of that section. Adopting the shorthand
\[\sigma^i \equiv \sigma(v^i), \quad (\sigma^\gamma)^i \equiv
\sigma^\gamma (v^i)\]
\[ {\beta}^{i}_{j} \equiv \beta ({t^i}_j),\qquad {(\beta^\gamma)^i}_j
\equiv \beta^\gamma ({t^i}_j),\qquad {F^i}_j \equiv F({t^i}_j), \quad
{\gamma^i}_j \equiv \gamma ({t^i}_j)\]
we have the following formulae:
\[(\sigma^\gamma)^i = {\gamma ^i}_j \sigma^j\]
\[{(\beta^\gamma)^i}_j = {\gamma^i}_k{\beta^k}_l {(\gamma^{-1})^l}_j
+ {\gamma^i}_kd{(\gamma^{-1})^k}_j\]
\[\nabla \sigma ^i = d\sigma ^i + {\beta ^i}_j \sigma^j\]
\[{F^i}_j = d {\beta^i}_j + {\beta ^i}_k {\beta ^k}_j\]
\[\nabla^2 \sigma ^i = {F^i}_j \sigma^j\]
\[d {F^i}_j + {\beta^i}_k {F^k}_j - {F^i}_k{\beta ^k}_j =0.\]
This describes a matrix example of our quantum-group gauge theory in
the left-handed conventions that appeared in the main part of
Section~3.

Now consider $V = \overline{Z}(R)$, where $\overline Z(R)$ is an algebra
generated by the set $\vecw=(w_i)_{i=1}^{n}$ modulo the following relations and
right $A(R)$-coaction
\[w_1w_2R = \lambda w_2w_1,\qquad  \rho_R w_i = w_j \otimes {t^j}_i\]
where, as previously, $\lambda \in k^*$. One can easily check that
$\overline Z(R)$ is right $A(R)^{\rm op}$-comodule algebra with $\rho_R$
as stated.  We suppose it quotients aslo to a coaction of $A$.

If $B$ is any algebra with unit then $E(B,\overline Z(R),A)$ is a
trivial right quantum vector bundle. Adopting the shorthand
\[\sigma_i \equiv \sigma(w_i), \quad (\sigma^\gamma)_i \equiv
\sigma^\gamma (w_i)\]
we now have the following formulae:
\[(\sigma^\gamma)_j = \sigma_i{\gamma ^i}_j \]
\[{(\beta^\gamma)^i}_j = {(\gamma^{-1})^i}_k{\beta^k}_l {\gamma^l}_j
+ {(\gamma^{-1})^i}_kd{\gamma^k}_j\]
\[\nabla \sigma_j = d\sigma_j - \sigma_i{\beta ^i}_j \]
\[{F^i}_j = d {\beta^i}_j + {\beta ^i}_k {\beta ^k}_j\]
\[\nabla^2 \sigma_ji = -\sigma_i{F^i}_j \]
\[d {F^i}_j + {\beta^i}_k {F^k}_j - {F^i}_k{\beta ^k}_j =0.\]
This describes a matrix example of our quantum-group gauge theory in
the right-handed conventions that appeared at the end of Section~3.

\end{document}